\documentclass[sigconf, nonacm]{acmart}
\usepackage{subfigure}

\AtBeginDocument{%
  \providecommand\BibTeX{{%
    \normalfont B\kern-0.5em{\scshape i\kern-0.25em b}\kern-0.8em\TeX}}}

\setcopyright{acmlicensed}
\copyrightyear{2018}
\acmYear{2018}
\acmDOI{XXXXXXX.XXXXXXX}

\acmConference[Conference acronym 'XX]{Make sure to enter the correct
  conference title from your rights confirmation emai}{June 03--05,
  2018}{Woodstock, NY}
\acmISBN{978-1-4503-XXXX-X/18/06}

\usepackage{algorithm}
\usepackage{algorithmicx}
\usepackage{algpseudocode}
\usepackage{multirow}
\usepackage{amsmath}
\usepackage{stfloats}

\begin{document}

\title{Journey into SPH Simulation: \\ A Comprehensive Framework and Showcase}


\author{Haofeng Huang}
\affiliation{%
  \institution{Institute for Interdisciplinary Information Sciences, \\Tsinghua University}
  \city{Beijing}
  \country{China}}
\email{huanghf22@mails.tsinghua.edu.cn}

\author{Li Yi}
\affiliation{%
  \institution{Institute for Interdisciplinary Information Sciences, \\Tsinghua University}
  \city{Beijing}
  \country{China}}
\email{ericyi@mail.tsinghua.edu.cn}

\renewcommand{\shortauthors}{Huang}

\begin{abstract}
This report presents the development and results of an advanced SPH (Smoothed Particle Hydrodynamics) simulation framework, designed for high fidelity fluid dynamics modeling. Our framework, accessible at \textcolor{blue}{\url{https://github.com/jason-huang03/SPH_Project}}, integrates various SPH algorithms including WCSPH, PCISPH, and DFSPH, alongside techniques for rigid-fluid coupling and high viscosity fluid simulations. Leveraging the computational power of CUDA and the versatility of Taichi, the framework excels in handling large-scale simulations with millions of particles. We demonstrate the capability of our framework through a series of simulations showcasing rigid-fluid coupling, high viscosity fluids, and large-scale fluid dynamics. Furthermore, a detailed performance analysis reveals CUDA’s superior efficiency across different hardware platforms. This work is an exploraion into modern SPH simulation techniques, showcasing their practical implementation and capabilities.
\end{abstract}

\begin{teaserfigure}
  \includegraphics[width=\textwidth]{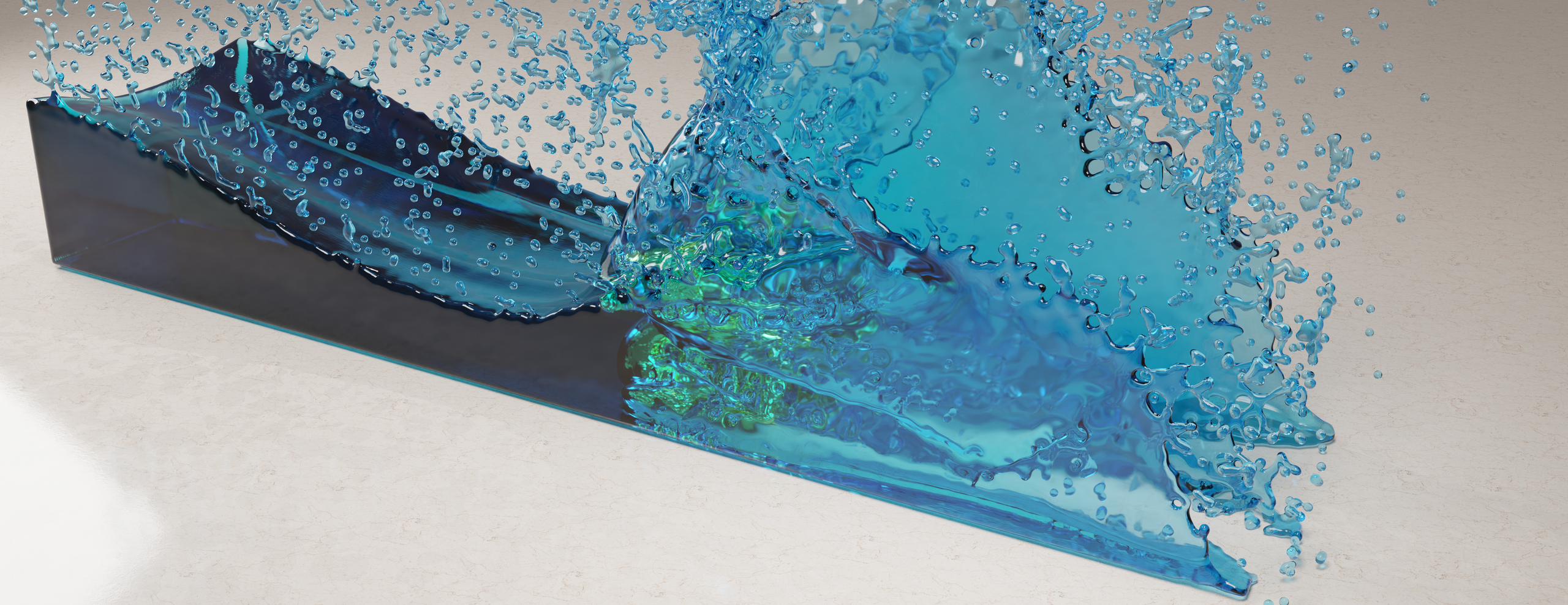}
  \caption{A glimpse into the dynamic world of our SPH simulations framework. This image captures the intricate beauty of 1.23M fluid particles splashing against rigid objects, showcasing the capability of our framework to produce detailed and realistic fluid simulation.}
  \Description{This teaser image presents a captivating snapshot from our SPH framework, hinting at the immense possibilities within. It features a moment frozen in time, where the fluid's intricate movements and interactions hint at the complex and expansive simulations our framework is capable of. The focus on specific fluid dynamics details invites the viewer to ponder the scale and depth of the simulations, showcasing the framework's prowess in a visually striking and thought-provoking manner.}
  \label{fig:teaser}
\end{teaserfigure}

\maketitle

\section{Introduction}
In the realm of computer graphics, the simulation of fluid dynamics has long been a topic of both challenge and fascination. The ability to realistically simulate fluid behaviors is crucial in various applications, ranging from entertainment to scientific visualization. However, achieving a balance between computational efficiency and visual realism remains a significant challenge.

The art of fluid simulation is a diverse landscape, characterized by different computational perspectives. Primarily, these include the Lagrangian view, the Eulerian view, and hybrid methods, each offering unique insights and approaches to fluid dynamics. The Lagrangian view, in particular, focuses on tracking fluid particles over time, providing a detailed and individualized understanding of fluid motion. In contrast, the Eulerian view concentrates on fluid properties at specific locations in space, offering a global perspective of fluid flow. Hybrid methods combine aspects of both, striving to leverage the strengths of each approach for more comprehensive simulations.

SPH (Smoothed Particle Hydrodynamics) stands out in the Lagrangian method. It models fluids as a collection of discrete elements, each representing a small volume of fluid. SPH not only intuitively simulates fluid behavior but also adeptly manages complex and free-surface flows, allowing for a high degree of flexibility and realism in the simulation.

Recent advancements in SPH have been impressive, expanding its capabilities to model not only fluid dynamics but also interactions among rigid and elastic bodies, as well as multi-phase fluids, all within a unified framework. This progress has substantially broadened the scope of SPH, enabling it to simulate a wider array of physical phenomena with enhanced realism and efficiency. These developments have also significantly refined SPH's capacity to handle a variety of fluid viscosities, making it an adaptive tool for simulating a spectrum of scenarios from flowing water to molten lava.

In this project, we delve deeply into the world of SPH by implementing several distinct SPH fluid simulation algorithms. These algorithms highlight the versatility of SPH, demonstrating its ability to simulate a wide range of fluid behaviors. From modeling basic liquid dynamics to capturing more complex fluid interactions, our implementations show the extensive potential of SPH in various scenarios.

An important part of our project is the integration of a rigid-fluid coupling method. This is crucial for creating realistic interactions between solid objects and fluids.  This allows us to accurately depict scenarios where solids and fluids interact (see Figure~\ref{fig:coupling}, Figure~\ref{fig:large_scale}), like objects moving through water or liquids spilling over solid surfaces.

Another focus of our project is on exploring the simulation of high viscosity fluids within the SPH framework, a complex aspect of fluid dynamics drawing increasing attention in recent years. High viscosity scenarios, such as the flow of molten lava or the slow drift of thick oils, present unique challenges in accurately depicting their distinct properties and behaviors. In our project we have carefully integrated a state-of-the-art method for high viscosity fluid simulation into our framework, significantly enhancing the realism in our fluid dynamics simulations and broadening the scope of what can be achieved under our framework (see Figure~\ref{fig:high_viscosity},  Figure~\ref{fig:buckling}, Figure~\ref{fig:coiling}).

In summary, our project demonstrates a comprehensive application of SPH-based fluid dynamics simulations in computer graphics. We successfully implemented several SPH algorithms,  along with a robust rigid-fluid coupling technique and a state-of-the-art high viscosity fluid simulation method.  Key to our project's success is the integration of GPU acceleration for enhanced computational efficiency, complemented by the application of established methods for efficiently solving linear systems. This combination enables us to scale our simulations up to millions of particles, significantly broadening the scope and realism of our simulations. Additionally, the flexibility of free scene setting in our framework and the use of an industrial-grade renderer have been instrumental in elevating the quality and realism of our outputs, allowing for more visually impressive representations of fluid dynamics.

\section{Related Work}
This section is organized as follows: We begin by briefly discussing popular SPH pressure solvers, then review modern methods for rigid-fluid coupling, and conclude with an examination of different approaches to modeling viscosity within the SPH framework.

\subsection{SPH Pressure Solver}
Earlier pressure solvers in SPH utilized the Equation of State (EOS) approach, introduced to the computer graphics community in \cite{eos_first}. The method was later extended for spatial adaptivity in \cite{adaptive}. Weakly Compressible SPH (WCSPH) \cite{wcsph} proposed an EOS method to limit compression using predetermined stiffness coefficients. However, achieving minimal density changes requires high stiffness coefficients, leading to stiff differential equations and limiting the maximum time step.

In contrast to the EOS solvers, there is another class of solvers that relies on Pressure Poisson Equation (PPE). These solvers typically advect velocity based on non-pressure forces and then iteratively refine velocity through pressure calculations based on PPE. Notable examples include Predictive-Corrective Incompressible SPH (PCISPH) \cite{pcisph}, and Divergence-Free SPH (DFSPH) \cite{dfsph} with DFSPH being state-of-the-art. While PPE solvers are computationally more intensive per step compared to EOS solvers, they can operate with larger time steps due to their more accurate pressure computation. Also, in an EOS solver the stiffness coefficient has to be manually specified while in most PPE solvers we only need to define the max error rate.

Besides these two classes of solvers there exist other kind of solvers that are closely related to a PPE solver. For example, Position Based Fluids (PBF) \cite{pbf} use position based dynamics \cite{pbd} to enforce constant density after advecting the particles with non-pressure forces. \cite{cf} enforces incompressibility and boundary conditions using holonomic kinematic constraints on the density.

\subsection{Rigid-Fluid Coupling in SPH}
Rigid-Fluid coupling under Lagrangian view has long been discussed. In this section we only focus on recent progress. \cite{akinci2012} samples rigid body as SPH particles and model the rigid-fluid interation as pressure forces. This formulation is particularly useful for incomplete neighborhoods and non-uniform boundary sampling, \textit{i.e.} one-layer-boundaries with particles spaced unevenly can be well handled. \cite{akinci2013} extends this concept to elastic solids. \cite{density_map} proposes density maps instead of boundary particles to compute the influence of a rigid body onto the fluid.  Based on \cite{density_map}, \cite{volume2019} make improvement on density map to make it better incorporate into SPH methods. In all these methods, the velocities of rigid bodies are considered to be constant when solving fluid dynamics.

In contrast, \cite{interlinked} shows that it is possible to model rigid contact under SPH framework and propose a SPH rigid solver. It interconnects two SPH pressure solvers to simultaneously resolve fluid-rigid and rigid-rigid contact, instead of fixing the velocities of rigid bodies in the fluid solver. This is quite different from previous methods, offering improved results over traditional methods.

\subsection{Viscosity Calculation in SPH}
The standard method for computing viscosity forces, suitable for low-viscosity fluids, is proposed in \cite{mon05}, where the Laplacian of velocity is explicitly calculated. 

In recent years the simulation of highly viscous fluids has become popular. Therefore, implicit mdthods are required for a stable simulation. \cite{tdf15} models viscosity based on the strain-rate and use a backward Euler scheme for time integration. \cite{pict15} proposes an implicit solver that decomposes the velocity gradient, projects the field onto a shear rate reduced state and reconstructs the velocity field. \cite{pt16} extended this approach by vorticity difusion to improve the rotational motion. \cite{bk17} proposes a constraint based formulation where it projects the strain rate tensor onto a reduced state. In contrast to the previous approaches, \cite{wkbb18} introduces an implicit viscosity solver based on the Laplacian of the velocity field instead of using the strain rate.

\begin{figure*}[t]
    \includegraphics[width=\textwidth]{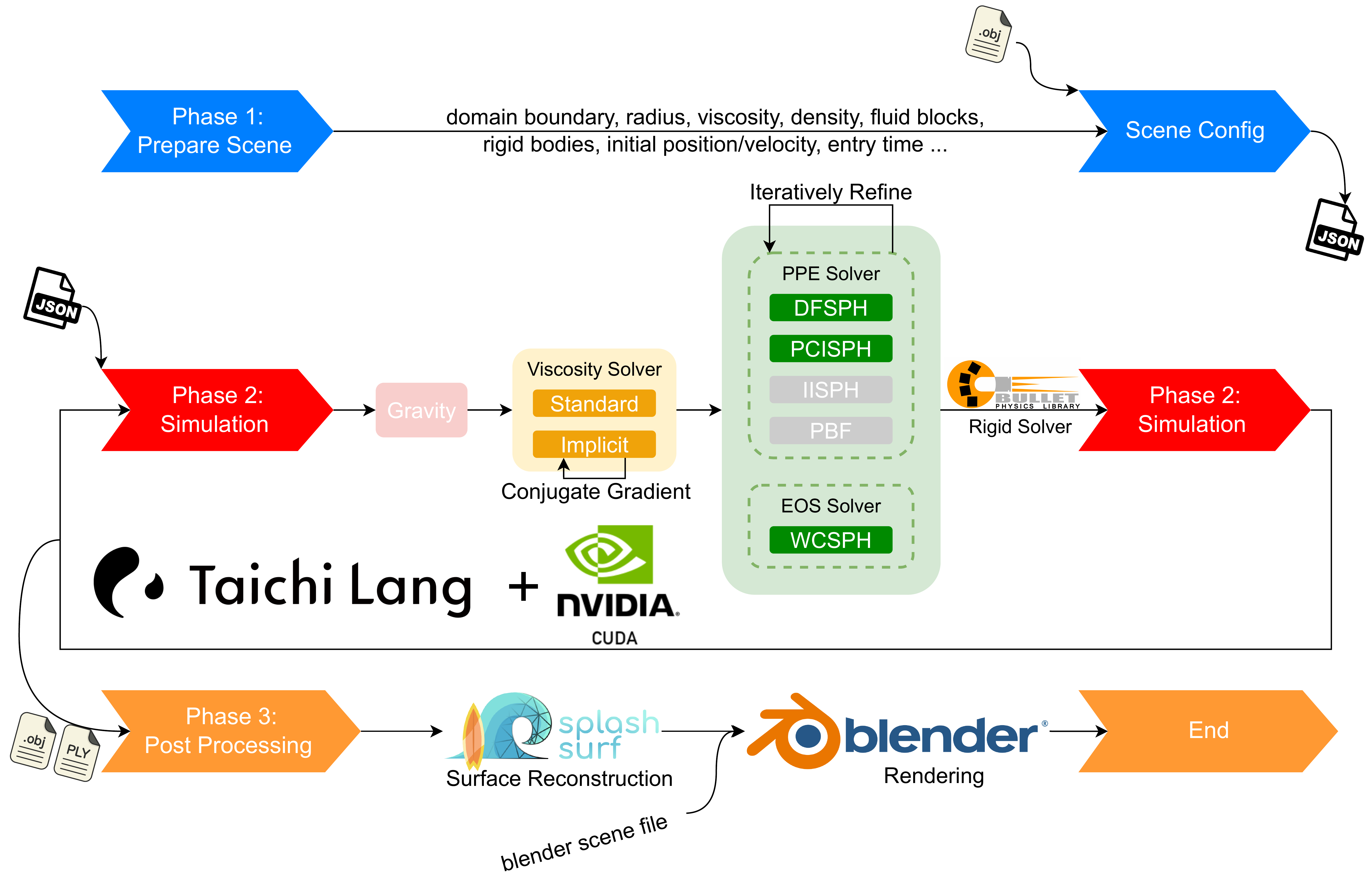}
\caption{
Detailed workflow of our SPH simulation framework.  This diagram illustrates the step-by-step process of our SPH simulation pipeline. It begins with the "Prepare Scene" phase, where users define fluid and rigid body properties, along with simulation parameters. The "Simulation" phase is next, showcasing the sequential application of gravitational, viscous, and pressure forces, followed by the dynamic updates of fluid and rigid bodies, incorporating various solvers for viscosity and pressure management. The final "Post-Processing" phase involves surface reconstruction of fluid particles using SplashSurf and rendering the scene in Blender, utilizing Blender's rich community resources for scene creation.
}
\label{fig:framework_overview}
\end{figure*}

\section{Method}
\subsection{Fluid Dynamics}
In fluid dynamics, the behavior of fluids is primarily governed by two fundamental equations. The first is the continuity equation, which relates to the evolution of an object's mass density \(\rho\) over time:
\begin{equation}
    \frac{D \rho}{D t} = - \rho (\nabla \cdot \mathbf{v})
\end{equation}
where \(\frac{D (\cdot) }{D t}\) denotes the material derivative. In the context of Smoothed Particle Hydrodynamics (SPH), which often deals with incompressible materials, this equation simplifies to:
\begin{equation}
	\frac{D \rho}{D t} = 0 \quad \Leftrightarrow \quad \nabla \cdot \mathbf{v} = 0 \label{eq:incompressibility_constraint}
\end{equation}
This constraint reflects the incompressibility of the material, meaning the density remains constant over time.

The second critical equation in fluid dynamics is the Navier-Stokes equation, serving as the equation of motion for incompressible flow:
\begin{equation}
        \rho  \frac{D \mathbf{v} }{D t}  = - \nabla p + \mu \nabla ^{2} \mathbf{v}  + \mathbf{f} _{\text{ext} }\label{eq:navier_stokes}
\end{equation}
where \(p\) denotes the pressure, \(\mu\) is the viscosity coefficient and \(\mathbf{f}_{\text{ext}}\) is the body force per volume. This equation is fundamental in describing how fluid velocity \(\mathbf{v}\) changes over time under the influence of various forces.

SPH algorithms in essence solve these two equations with numerical methods under additional boundary constraints.

\subsection{SPH Foundations}
The high level idea of SPH is to use particles "carrying" samples of field quantities, and a kernel function \(W: \mathbb{R}^d \times \mathbb{R}^+ \to \mathbb{R}\), to approximate continous fields. Intuitively we can imagine each particle as a small parcel of water carrying quantities (although strictly not the case). Actually a set of SPH particles is a spatial function descretization. For detailed formulation of SPH, we refer the reader to \cite{tutorial2019} and \cite{tutorial2022}. For now, we suppose each particle has mass \(m_i\), position \(\mathbf{x}_i\), velocity \(\mathbf{v}_i\), density \(\rho_i\) and other quantities. The mass is fixed for each particle. 

The kernel function \(W\) plays a crucial role in the spatial discretization process. It serves the purpose of smoothing in spatial discretization. Typically, the \(W\) has a support radius \(h\), outside of which it becomes negligible or vanishes. We define \(W_{ij} = W(\mathbf{x}_i - \mathbf{x}_j, h)\) for practical computations. With this, the density can be derived from mass as
\begin{equation}
    \rho_i = \sum_{j} m_j W_{ij} \label{eq:density}
\end{equation}

For the gradient of a scalar quantity \(A_i\) (\textit{e.g.} gradient of density), we use the formula
\begin{equation}
    \nabla A_i = \rho_i \sum_{j} m_j\left( \frac{A_i}{\rho_i ^2} + \frac{A_j}{\rho_j ^2} \right)\label{eq:gradient}
\end{equation}

For the divergence of vector quantity \(\mathbf{A}_i\) (\textit{e.g.} devergence of velocity), we use the formula
\begin{equation}
        \nabla \cdot \mathbf{A} _{i} = \sum_{j}^{} \frac{m_j}{\rho  _{j}} (\mathbf{A} _j - \mathbf{A}_i)  \cdot \nabla W_{ij}
\end{equation}


\subsection{General SPH Algorithm}
The essence of SPH algorithm lies in  solving the Navier-Stokes equation (Eq.~\eqref{eq:navier_stokes}) under the incompressibility condition (Eq.~\eqref{eq:incompressibility_constraint}). Modern SPH algorithms commonly adopt the concept of  \emph{operator splitting}. This technique decomposes the complex partial differential equation into a sequence of simpler subproblems, each of which can be tackled with specialized solution techniques. 

A general pipeline for a single iteration in SPH algorithms is shown in Algorithm~\ref{alg:general_algorithm}. The Navier-Stokes equation is splitted into two primary components. 

Initially, we address the component excluding pressure forces. In the simplest scenarios, the only non-pressure force considered is gravity, although other forces like viscosity and surface tension can also be included. The handling of viscosity forces is further detailed in Section~\ref{sec:viscosity}. Here, non-pressure accelerations \(\mathbf{a}^{\text{nonp}}\) are calculated, and the a predicted velocity is computed with \(\mathbf{v} ^* = \mathbf{v} + \Delta t \mathbf{a}^{\text{nonp}} \).

Following this, the algorithm utilizes the velocities \(\mathbf{v}^*\) obtained from the first component to address the pressure-related component. The various approaches to solving the pressure equation are discussed in Section~\ref{sec:pressure_solver}. Through this process, the pressure \(p\) is determined, and pressure accelerations can be computed with \(\mathbf{a}^{\text{p}} = - \frac{1}{\rho} \nabla p\). Following Eq.~\eqref{eq:gradient}, we can get
\begin{equation}
    \mathbf{a}_i ^{\text{p}} = - \sum _j m_j \left( \frac{p_i}{\rho_i ^ 2} + \frac{p_j}{\rho _j ^ 2} \right) \nabla W _{ij}\label{eq:pressure_acceleration}
\end{equation}

Finally, the updated velocities and positions of the particles are computed using a symmetric Euler integration method.

\begin{algorithm}
    \caption{Single Iteration for General SPH Algorithm}
    \label{alg:general_algorithm}
    \begin{algorithmic}[1]
        \State         calculate non-pressure force acceleration \(\mathbf{a}^{\text{nonp}}\), which may include gravity, viscosity, surface tension and so on
        \State \(\mathbf{v} ^* = \mathbf{v} + \Delta t \mathbf{a}^{\text{nonp}}\)
        \State solve pressure \(p\) from \(\mathbf{v} ^*\)  (enforcing \(\frac{D \rho}{D t} = 0\))
        \State get pressure force acceleration by \(\mathbf{a}^{\text{p}} = - \frac{1}{\rho} \nabla p\)
        \State \(\mathbf{v}  = \mathbf{v} ^{*} + \Delta  t \mathbf{a} ^{\text{p} }\)
        \State \(\mathbf{x}  = \mathbf{x} + \Delta t \mathbf{v} \)

    \end{algorithmic}  
\end{algorithm}

\subsection{Pressure Solver}
\label{sec:pressure_solver}
Solving the pressure is the core of every SPH algorithm. The pressure field plays a crucial role in maintaining incompressibility and preserving the fluid's volume. It achieves this by exerting pressure acceleration \(\mathbf{a} ^{\text{p}} = - \frac{1}{\rho} \nabla p\). In this section, we delves into two primary categories of pressure solvers, each with its unique approach and characteristics. Additionally, it's important to note the existence of certain solvers that handle fluid dynamics without explicitly dealing with pressure, such as Position Based Fluids (PBF) \cite{pbf}.

\subsubsection{EOS Pressure Solver}
EOS (equation of state) solvers directly use current density deviation to compute pressure. This deviation is typically expressed as a quotient or a difference relative to the rest density. A commonly used formulation is \(p_i = k \left(  \left(\frac{\rho_i}{\rho^0}\right)^{\gamma} -1\right)\), where \(k\) is the stiffness coefficient 
 and \(\gamma\) is an exponent that defines the compressibility of the fluid. Such algorithms, exemplified by the well-known Weakly Compressible SPH (WCSPH) \cite{wcsph}, are often labeled as "weakly compressible", because they do not enforce zero density deviation directly.
 
 One of the main advantages of EOS solvers is their simplicity and ease of implementation. This makes them an ideal starting point for many who are developing an SPH fluid simulator for the first time. However, a notable drawback of this approach is the potential for noticeable visual artifacts. These artifacts arise because in EOS solvers, the pressure is computed based on density deviations rather than velocity fields, which can cause non-zero divergence in the fluid's velocity (thus leading to non-constant density). Despite this, the practicality and straightforward nature of EOS solvers make them a popular choice, and we have implemented the WCSPH algorithm in our project to demonstrate these principles.

\subsubsection{PPE Pressure Solver}
The Pressure Poisson Equation (PPE) solvers operate on a principle where the calculated pressure accelerations lead to velocity changes, subsequently resulting in particle displacements that restore each particle to its rest density. PPE solvers solve a linear system to compute the respective pressure field. PPE solvers typically come in two variants, each with a distinct focus: one using density as the source term and the other using velocity.

For the density-based PPE solver, the goal is to maintain constant density after the application of pressure forces. This is mathematically expressed as
\begin{equation}
        \Delta t \nabla ^{2} p(t) = \frac{\rho ^{0} - \rho ^{*}}{\Delta t}\label{eq:density_source} 
\end{equation}
Here,  \(\rho ^*\) denotes the predicted density from \(\mathbf{v}^*\) calculated following the  continuity equation:
\begin{equation}
    \rho ^{*} = \rho (t) - \Delta t \rho (t) \nabla \cdot \mathbf{v} ^{*}
\end{equation}

Alternatively, the velocity-based PPE solver aims for zero divergence in velocity after pressure application. It is formulated as
\begin{equation}
    \Delta t \nabla ^{2}p(t) = \rho (t) \nabla \cdot \mathbf{v} ^{*}\label{eq:velocity_source}
\end{equation}

Modern PPE solvers typically works in an iterative manner to solve Eq.~\eqref{eq:density_source} or Eq.~\eqref{eq:velocity_source}. The general workflow of a single iteration in such a solver is outlined in Algorithm~\ref{alg:ppe_solver}. In this project, we have implemented Predictive-Corrective Incompressible SPH (PCISPH) \cite{pcisph} and Divergence-Free SPH (DFSPH) \cite{dfsph}. Among the various PPE solvers, DFSPH, which simultaneously uses Eq.~\eqref{eq:density_source} and Eq.\eqref{eq:velocity_source} to solve the dynamics, has the best overall performance.

\begin{algorithm}
    \caption{General Workflow of Iterative PPE Solver}
    \label{alg:ppe_solver}
    \begin{algorithmic}[1]
        \State get \(\mathbf{v}^ *\) derived from non-pressure forces
        \State initialize \(l=0, p ^l, \mathbf{v}^{*,l}\)
        \While{\(\mathit{err} > \eta\)}
            \State refine \(p ^{l+1} ( \mathbf{v}^{*, l}, p^l)\)
            \State update \( \mathbf{v} ^{*, l+1}(p  ^{l+1})\)
            \State compute \(\mathit{err}\) from density deviation or velocity divergence
        \EndWhile
    \end{algorithmic}  
\end{algorithm}

PPE solvers generally yield more accurate simulations, although they may require more computation time per step compared with EOS solvers. This trade-off is often offset by their enhanced stability, which allows for the use of larger time steps \(\Delta t\). In contrast to EOS solvers, where achieving a desired density deviation requires manually setting a stiffness coefficient \(k\), PPE solvers provide a more direct approach to controlling simulation accuracy. They do this by explicitly specifying the allowable maximum error, offering more precise management over the simulation's fidelity and outcomes.

\begin{figure}[h]
  \setlength{\tabcolsep}{2pt}
\begin{tabular}{ccc}
    \centering
    &\includegraphics[width=0.95\linewidth]{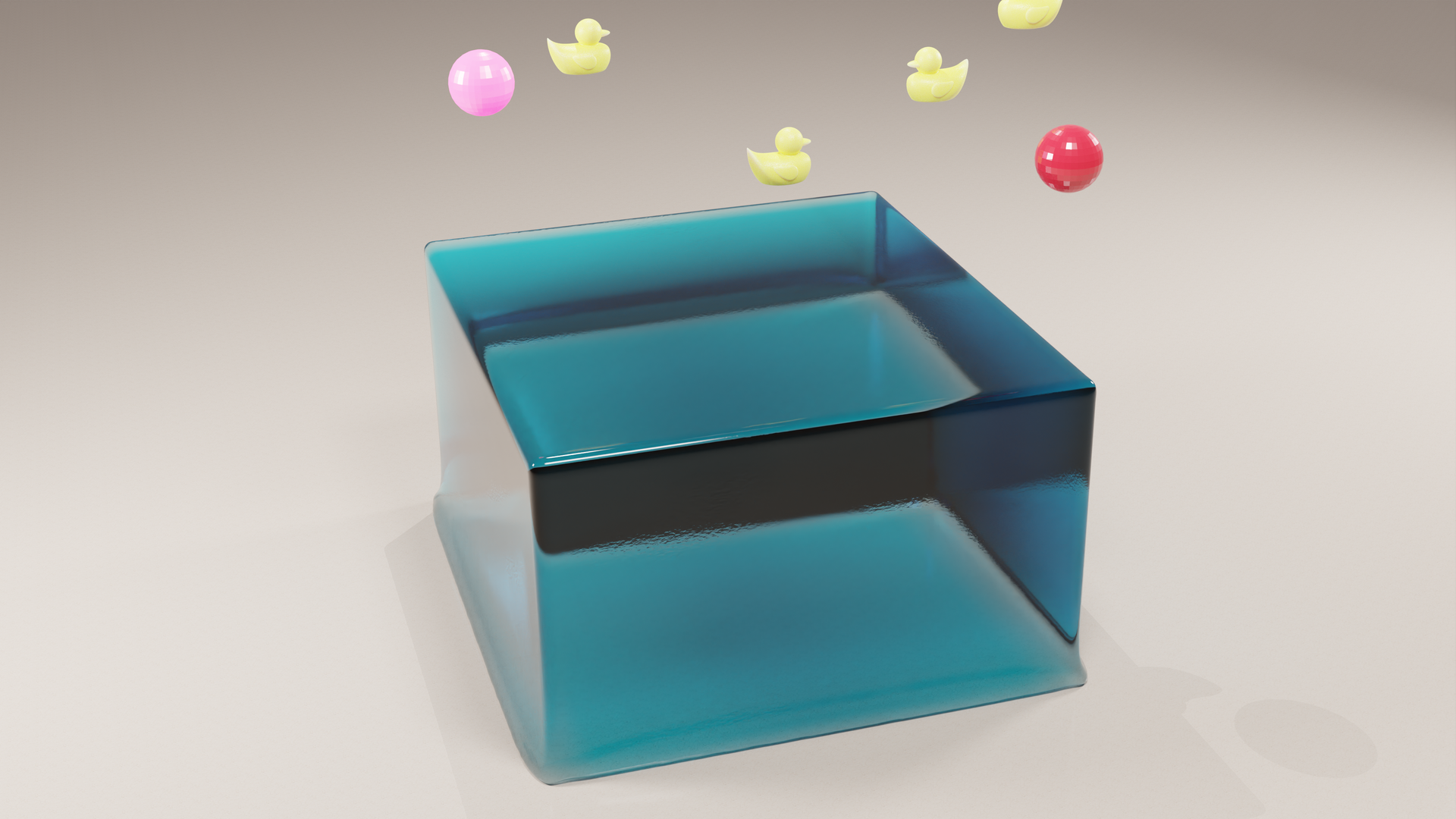} &  \\
    &\includegraphics[width=0.95\linewidth]{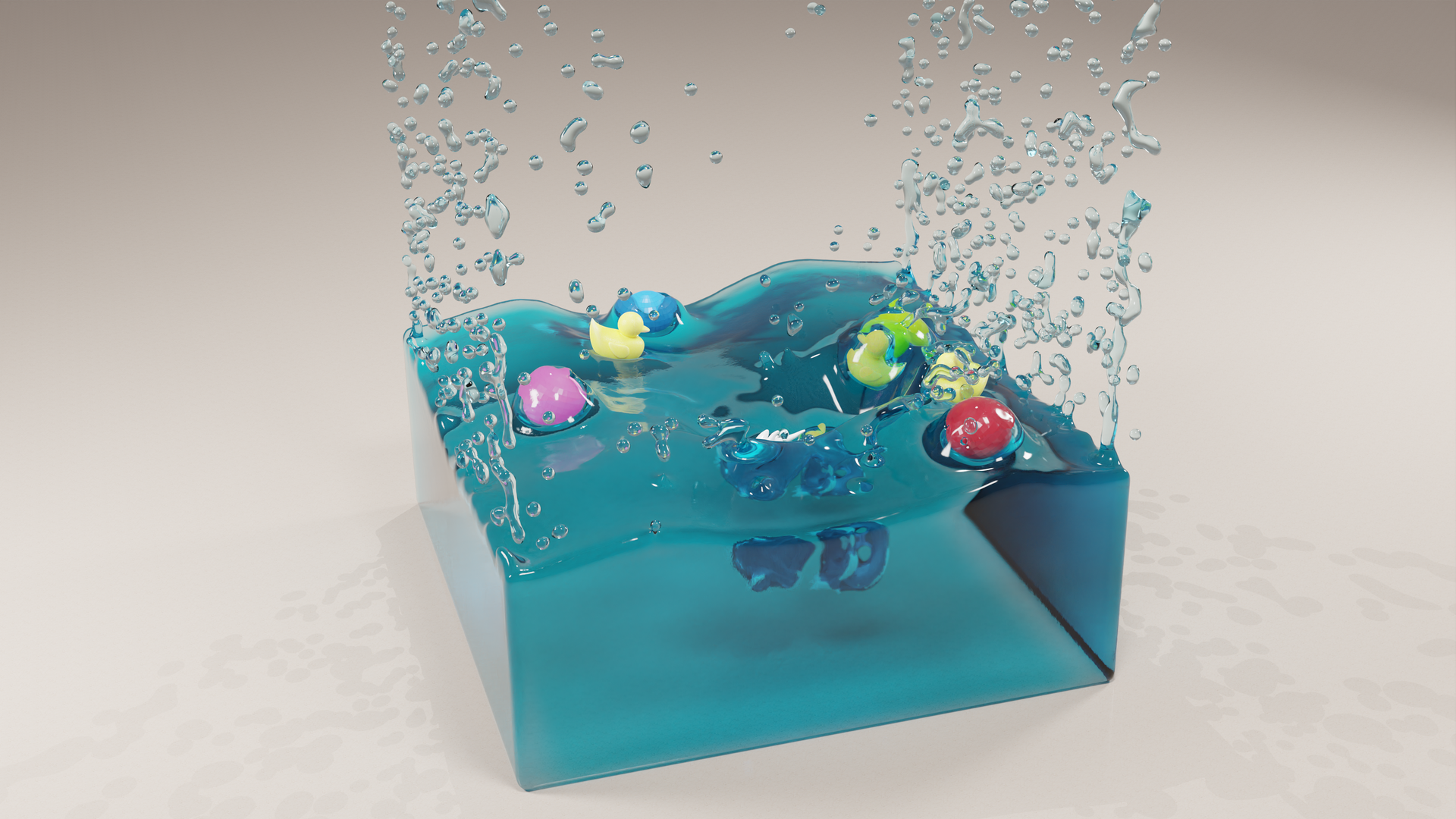} & \\
    &\includegraphics[width=0.95\linewidth]{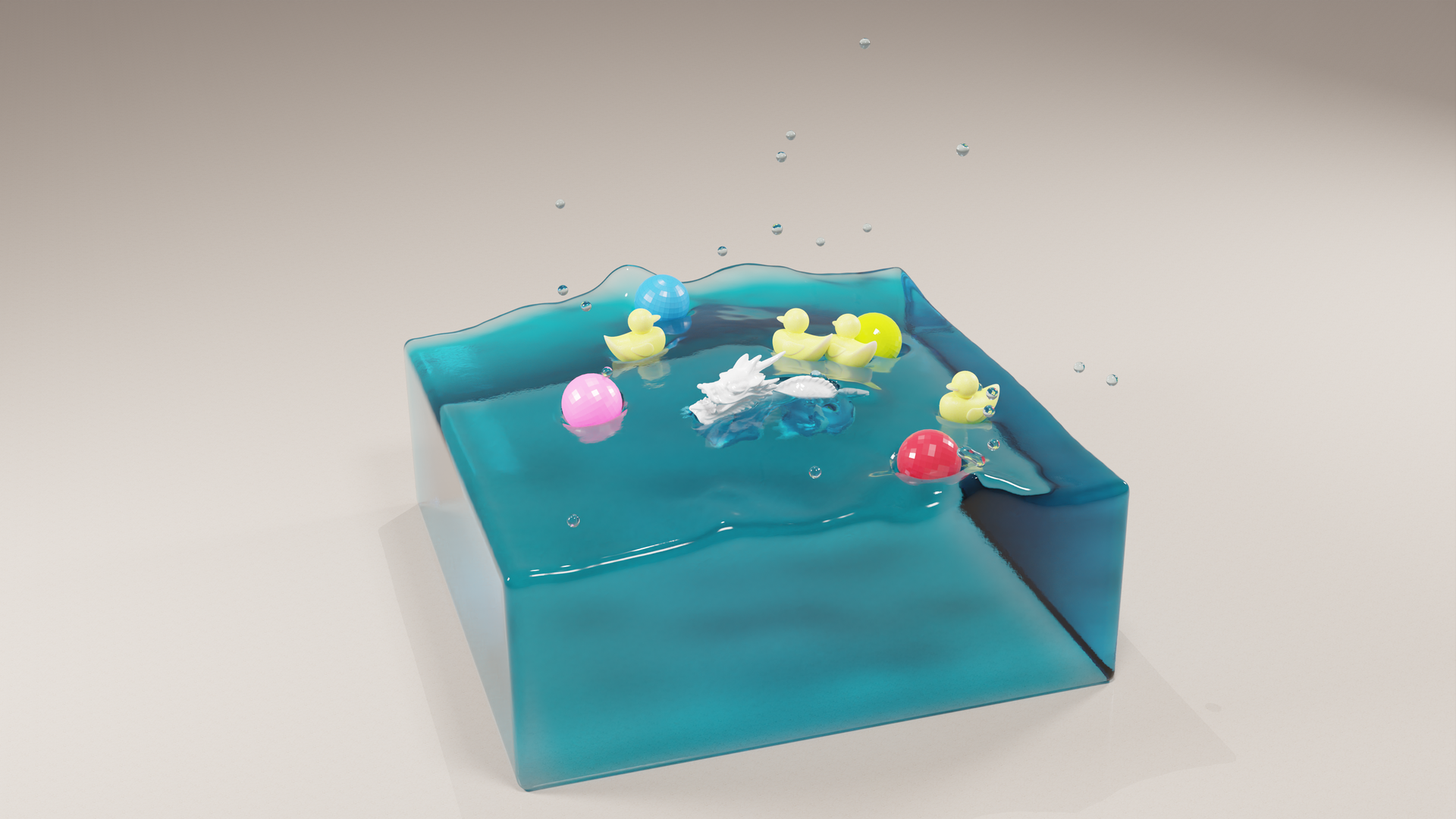} &
\end{tabular}
  \caption{Demonstration of Rigid-Fluid Coupling.}
  \label{fig:coupling}
  \Description{the complex interplay of forces at the interface. The image vividly demonstrates how the fluid conforms, flows around, or impacts the rigid bodies, showcasing the accuracy and realism with which our framework handles these interactions. The detail in the fluid's response to the presence and movement of the rigid objects highlights the sophistication of the simulation techniques used.

}
\end{figure}

\subsection{Rigid-Fluid Coupling}
\label{sec:coupling}
In our approach to tackling the complexities of rigid-fluid coupling within our SPH simulations, we've adopted the method proposed in \cite{akinci2012}. This method use boundary particles to sample the surface of rigid objects. One of the key strengths of this method is its versatility and robustness, accommodating various sampling techniques for the rigid boundaries. These boundaries can be represented with particles that are either evenly spaced or unevenly spaced, and they can consist of either a single layer or multiple layers of particles.

For any rigid particle \(b_i\), we can define its artificial mass as
\begin{equation}
\Psi _{b_i} (\rho _0) = \rho _0  \frac{1}{\sum_{k} W_{ik}}
\end{equation}
Here \(k\) denotes the rigid particle neighbors. 

Based on Eq.~\ref{eq:density}, this leads to the density of a fluid particle \(f_i\) being
\begin{equation}
\rho _{f_i} = \sum_{j} m_{f_j} W_{ij} + \sum _{k} \Psi_{b_k} (\rho_0) W_{ik}
\end{equation}
In this equation, \(j\) denotes the fluid particle neighbors. This formulation well handles the issue of density deficiency in fluid particles near boundaries, ensuring more accurate simulations of fluid-rigid interactions.

Inspired by Eq.~\eqref{eq:pressure_acceleration}, we write the pressure force applied from a rigid particle \(b_j\) to a fluid particle \(f_i\) as 
\begin{equation}
\mathbf{F}^{\text{p}}_{f_i \leftarrow b_j} = -m_{f_i} \Psi _{b_j} (\rho_0) \left( \frac{p_{f_i}}{\rho^2 _{f_i}}\right) \nabla W_{ij}\label{eq:rigid_pressure}
\end{equation}
The symmetric pressure force from a fluid particle to a rigid particle is
\begin{equation}
    \mathbf{F}^{\text{p}}_{b_j \leftarrow f_i} = - \mathbf{F}^{\text{p}}_{f_i \leftarrow b_j}\label{eq:rigid_pressure_symmetric}
\end{equation}
In Eq.~\eqref{eq:rigid_pressure} and Eq.~\eqref{eq:rigid_pressure_symmetric}, the idea is making use of a fluid particle's own pressure when computing the rigid-fluid interaction force. This formulation is quite robust and can handles rigid-fluid object interactions effectively (see Figure~\ref{fig:coupling}).

\subsection{Viscosity Solver}
\label{sec:viscosity}
In Navier-Stokes equation (Eq.~\eqref{eq:navier_stokes}) the viscous force is characterized by the Laplacian of the velocity field. Successfully estimating this Laplacian is key to accurately resolving the viscous forces in fluid dynamics simulations.

\subsubsection{Explicit Viscosity}
The standard approach, as proposed in \cite{mon05}, involves explicitly computing the Laplacian:
\begin{equation}
\label{eq:lap}
    \nabla ^2 \mathbf{v}_i = 2(d+2) \sum_{j} \frac{m_j}{\rho _j} \frac{\mathbf{v}_{ij} \cdot \mathbf{x}_{ij}}{\lVert \mathbf{x}_{ij} \rVert ^2 + 0.01 h^2} \nabla W_{ij}
\end{equation}
Here \(\mathbf{x}_{ij} = \mathbf{x}_i - \mathbf{x}_j\), \(\mathbf{v}_{ij} = \mathbf{v}_i - \mathbf{v}_j\), with \(d\)
 the the number of spatial dimensions and \(h\) the support radius of the kernel function \(W\). This simple formulation works quite well with low viscosity fluid, and we have implemented it within our framework.

\begin{figure*}
\setlength{\tabcolsep}{2pt}
\begin{tabular}{ccc}
    \centering
    \includegraphics[width=0.32\linewidth]{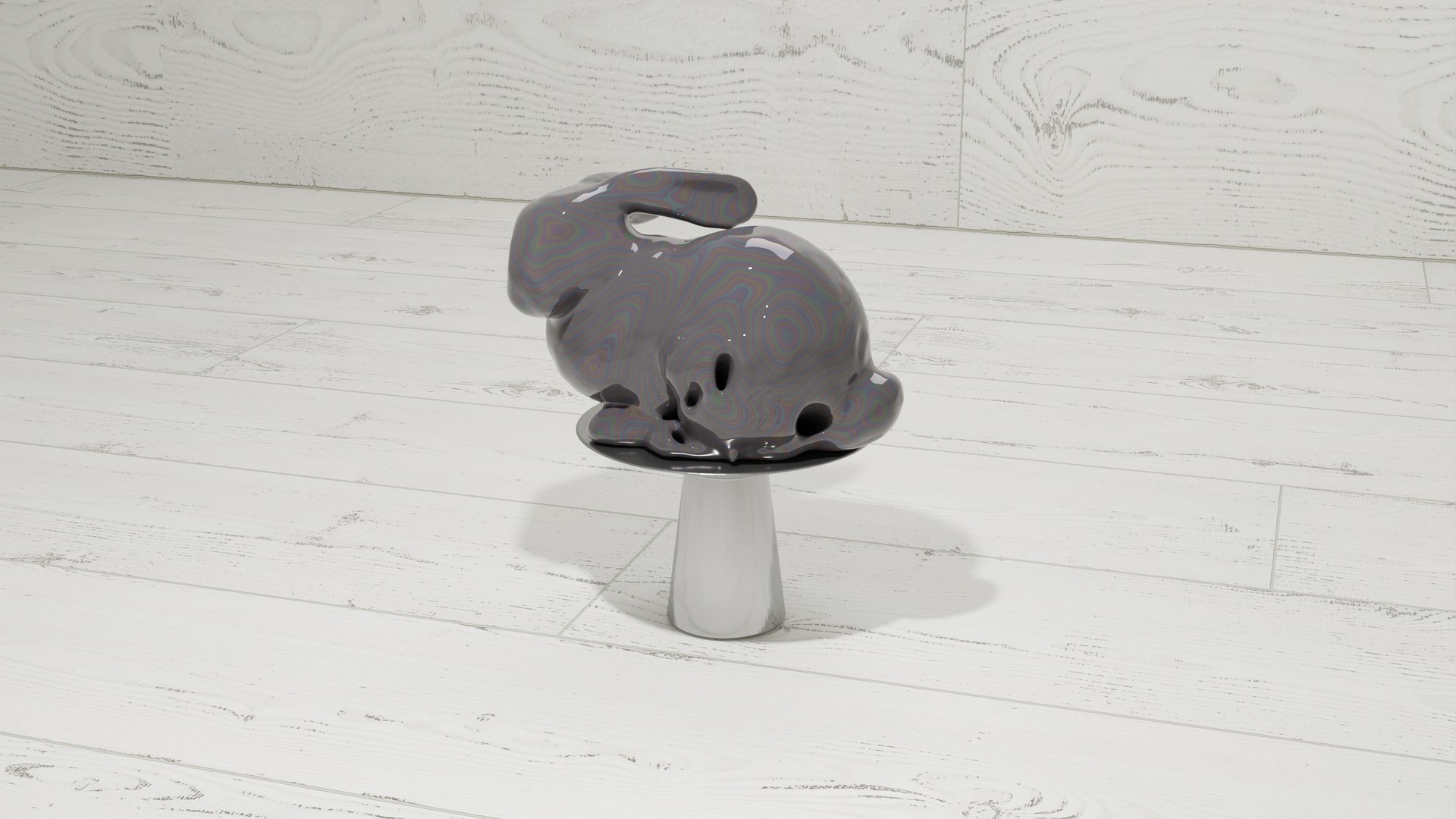} & 
    \includegraphics[width=0.32\linewidth]{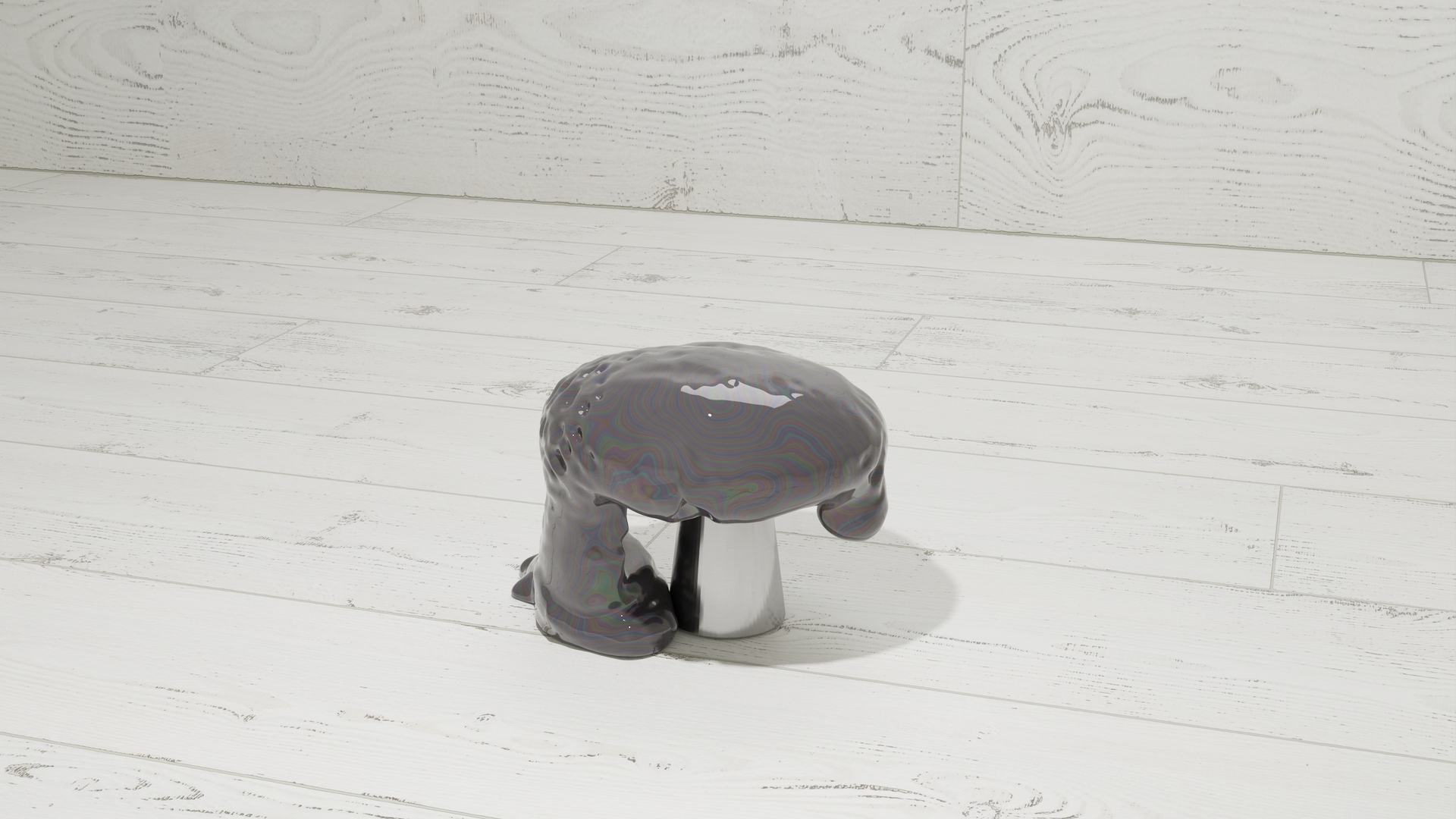} & 
    \includegraphics[width=0.32\linewidth]{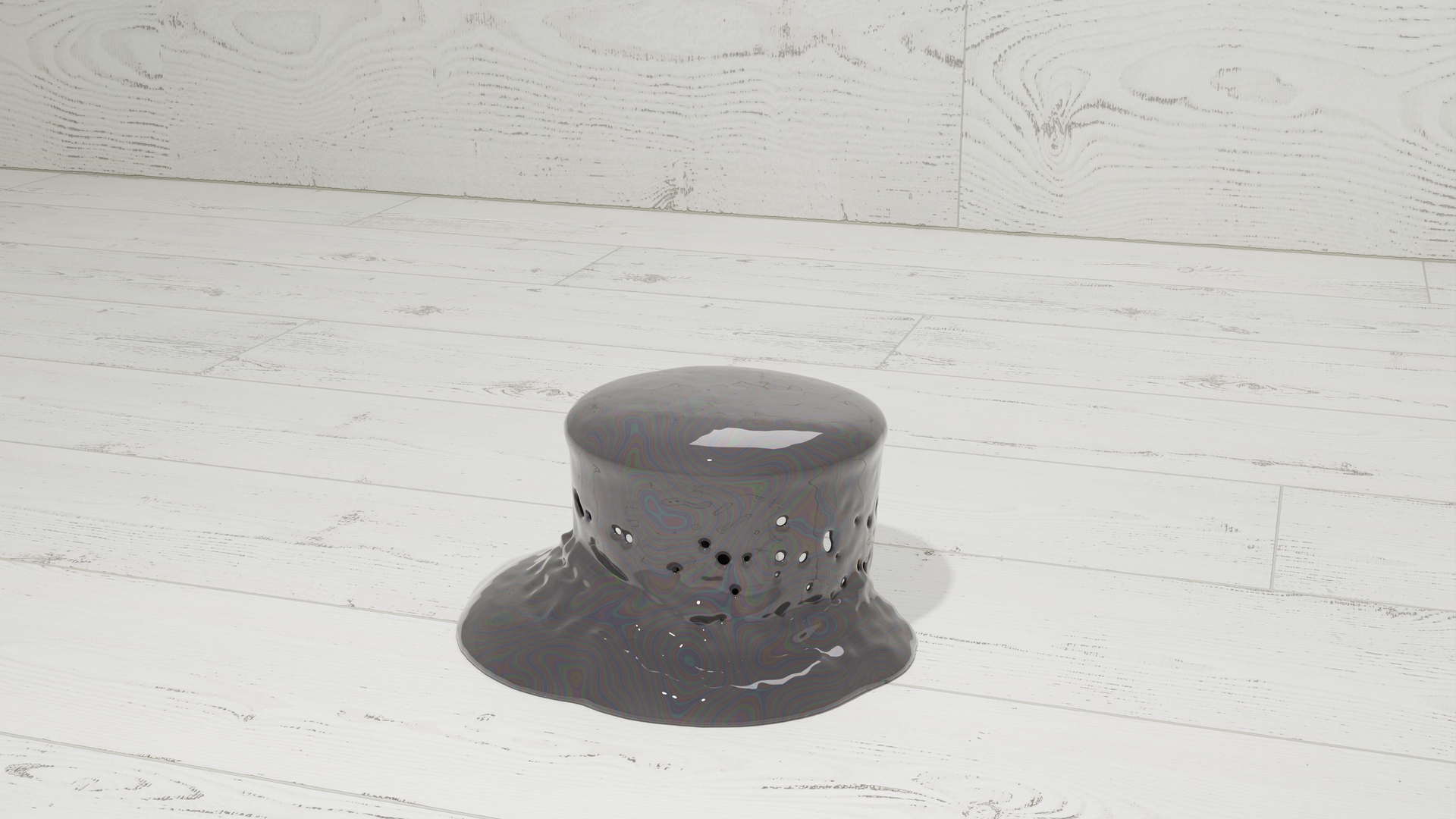} 
\end{tabular}
  \caption{Fluid with extremely high viscosity.}
  \label{fig:high_viscosity}
  \Description{This image vividly demonstrates the capabilities of our SPH simulation framework in modeling extremely high viscosity fluids. It depicts a fluid with a consistency resembling that of molasses or lava, capturing the slow, languid flow and the characteristic resistance to motion associated with highly viscous substances. The detail in the image highlights the fluid's gradual movement and the realistic deformation and texturing, showcasing the framework's precision and effectiveness in handling complex fluid dynamics.}
\end{figure*}
\subsubsection{Implicit Viscosity}

For fluids with high viscosity, explicit viscosity solvers may become unstable. To address this, implicit methods are preferred. In our framework, we have adopted the implicit viscosity method proposed in \cite{wkbb18}. This approach is particularly advantageous for handling high-viscosity fluids, as illustrated in our Figure~\ref{fig:high_viscosity}.

Let \(\Tilde{\mathbf{v}}\) represents the predicted velocity after applying all non-pressure forces except the viscous force. The explicit formulaition would be
\begin{equation}
    \mathbf{v} ^ * = \Tilde{\mathbf{v}} + \Delta t \frac{\mu}{\rho} \nabla ^ 2 \Tilde{\mathbf{v}}
\end{equation}
Here \(\mathbf{v}^*\) denotes the predicted velocity after applying all non-pressure forces. 

For stability, implicit integration is employed:
\begin{equation}
    \mathbf{v} ^* = \Tilde{\mathbf{v}} + \Delta t \frac{\mu}{\rho} \nabla ^2 \mathbf{v} ^ *
\end{equation}
Applying the formulation of Laplacian in Eq.~\eqref{eq:lap}, this leads to a linear system which can be expressed as
\begin{equation}
    (\mathbf{I} - \Delta t \mathbf{A})\mathbf{v}^* = \Tilde{\mathbf{v}}
\end{equation}
where the matrix \(\mathbf{A}\) consists of \(3 \times 3\) blocks
\begin{equation}
    \mathbf{A}_{ij} = -2(d+2) \frac{\mu \overline{m}_{ij}}{\rho _i \rho _j} \frac{\nabla W_{ij} \mathbf{x}_{ij} ^ T}{\lVert \mathbf{x}_{ij}\rVert + 0.01 h ^ 2}, \quad \mathbf{A}_{ii} = - \sum_{j} \mathbf{A}_{ij}
\end{equation}
Here \(\mathbf{A}_{ij}\) denotes the matrix block that corresponds to particles \(i\) and \(j\). The system is symmetric, therefore it allows for an efficient solution using the conjugate gradient method. Notably, this method can be implemented in a matrix-free manner, which is particularly advantageous for large-scale simulations where it is not possible to store all \(O(n^2)\) entries.

\section{Implementation Details}
In this project, we have crafted a highly versatile SPH simulation framework, designed to address a broad spectrum of fluid dynamics scenarios. An illustrative overview of our comprehensive pipeline is presented in Figure~\ref{fig:framework_overview}, capturing the essence of our methodical approach.

\subsection{Phase One: Configuration}
Initially, users prepare a configuration file to set up the simulation environment. This includes defining fluid and rigid bodies, where fluids can be added as blocks or through meshes, and rigid bodies are added via meshes. Users can also specify a range of simulation parameters like domain boundary, simulation method, particle radius, support radius, density, viscosity, initial position, velocity, rotation, and entry time.

\subsection{Phase Two: Core Simulation}
The simulation phase forms the cornerstone of our pipeline. Each iteration within this phase follows a systematic sequence of steps to simulate fluid dynamics accurately. Initially, gravity is applied to all particles. This is followed by the computation of viscous forces, and then the calculation of pressure forces. The iteration concludes with updating the dynamics of both the rigid bodies and the fluid particles. 

In tackling the various facets of fluid dynamics, we have integrated a range of specialized solvers into our framework. For handling viscosity, we employ both the standard explicit solver \cite{mon05} and an advanced implicit solver \cite{wkbb18}. The explicit solver offers a straightforward approach, while the implicit solver provides enhanced accuracy for more complex simulations, catering to different levels of complexity and accuracy requirements (details in Section~\ref{sec:viscosity}). The dynamics of pressure within the fluid are managed using a diverse set of algorithms, including Weakly Compressible SPH (WCSPH) \cite{wcsph}, Predictive-Corrective Incompressible SPH (PCISPH) \cite{pcisph}, and Divergence-Free SPH (DFSPH) \cite{dfsph},  each suited to different simulation needs (as elaborated in Section~\ref{sec:pressure_solver}) 

The interaction between the fluid and rigid bodies is modeled using the approach detailed in \cite{akinci2012} (discussed in Section~\ref{sec:coupling}), which allows for robust simulation of fluid-solid interactions. The dynamics of the rigid bodies are computed using the robust PyBullet physics engine \cite{bullet}. 

All these elements are integrated into a unified framework powered by Taichi \cite{hu2019taichi}\cite{hu2019difftaichi}\cite{hu2021quantaichi}, which allows parallel computation on CUDA. We also employ conjugate gradient (matrix free) and Jacobi methods for solving linear systems efficiently.

\subsection{Phase Three: Post-Processing}
After simulation, we move to post-processing. Surface reconstruction of fluid particles is done using SplashSurf \cite{splashsurf}, followed by rendering in Blender to visualize the results. In Blender, we leverage the extensive resources available within its community to build and enhance our scenes, ensuring that the final renderings are not only accurate but also visually compelling. 

\subsection{Additional Notes}
We have also experimented with implementing Implicit Incompressible (IISPH) \cite{iisph} and Position Based Fluids (PBF) \cite{pbf}. However, we faced challenges in correctly implementing IISPH and integrating PBF into our unified framework, indicating potential areas for future work.

\section{Results}
In this section, we present several compelling results generated using our framework. For all simulations, we consistently use the DFSPH solver because it offers the best overall performance, setting the particle radius to \(r=0.01\), supporting radius \(h=4r\) and \(\rho^0 = 1000 \, \text{kg/m}^3\). We does simulation on a single Nvidia A100 Tensor Core GPU. We render the result with Blender Cycle mode (device type OPTIX), 512 samples and OPTIX AI denoiser.

\begin{figure}[h]
  \setlength{\tabcolsep}{1pt}
\begin{tabular}{ccc}
    \centering
    &\includegraphics[width=0.96\linewidth]{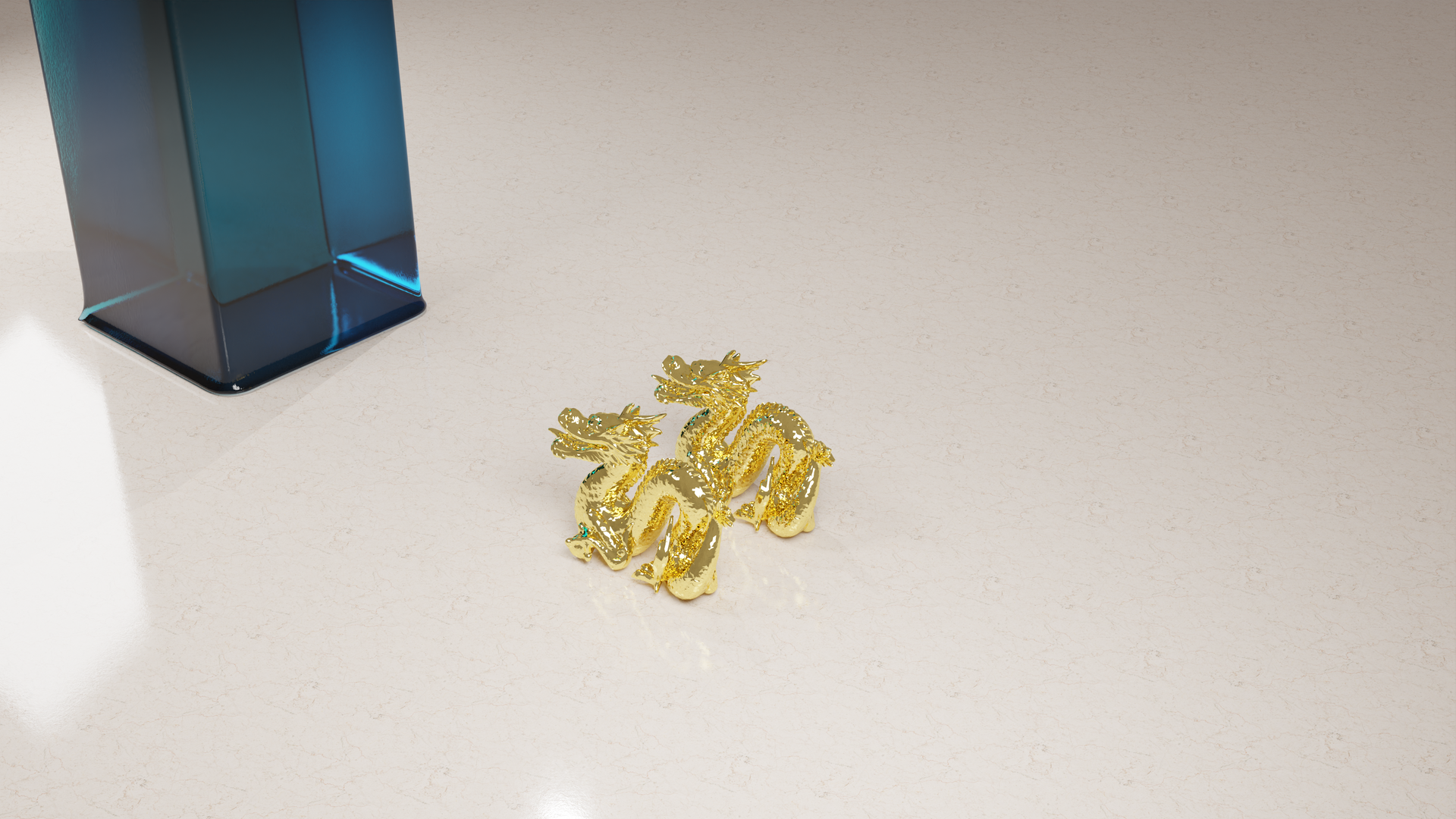} &  \\
    &\includegraphics[width=0.96\linewidth]{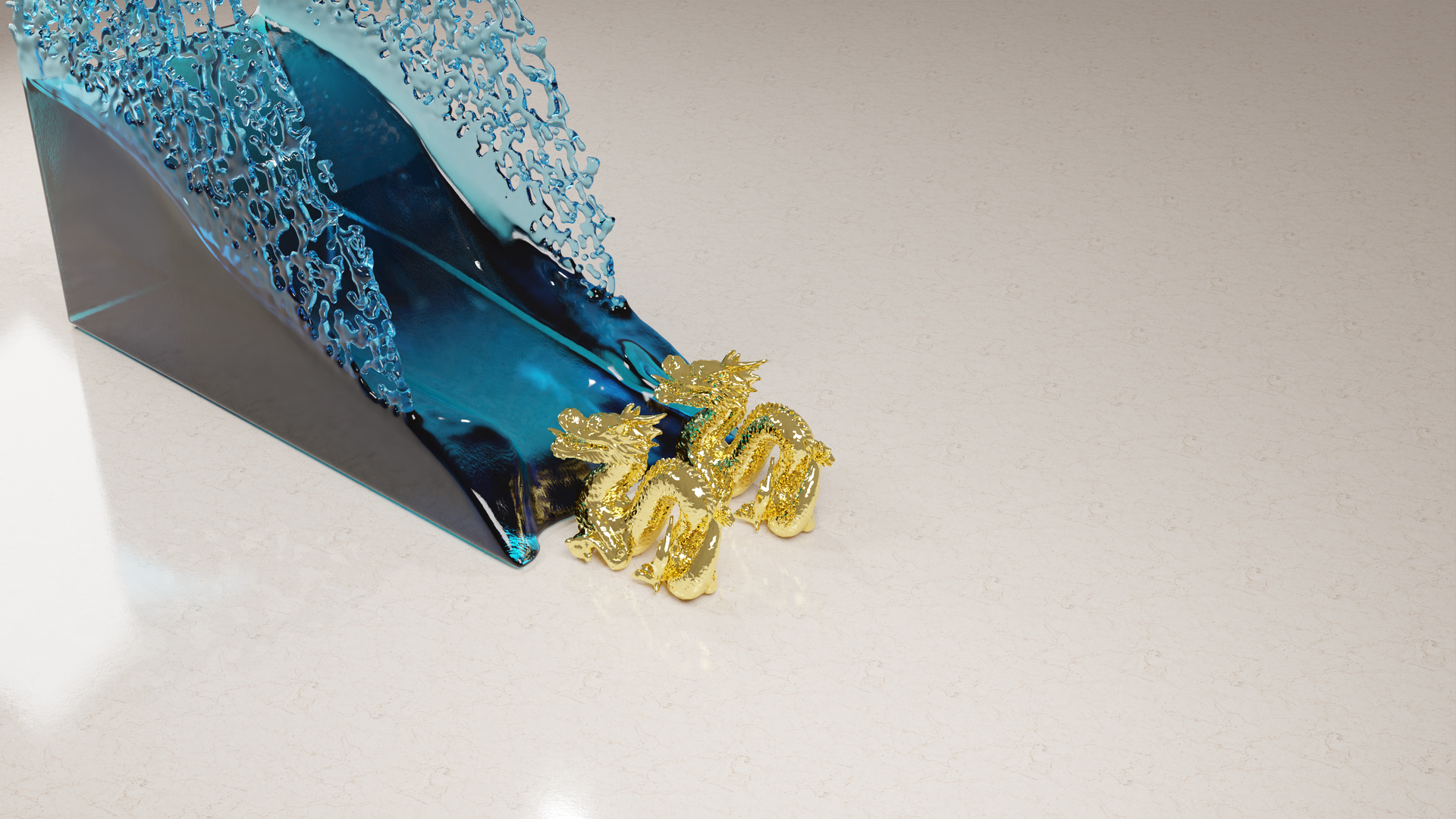} & \\
    &\includegraphics[width=0.96\linewidth]{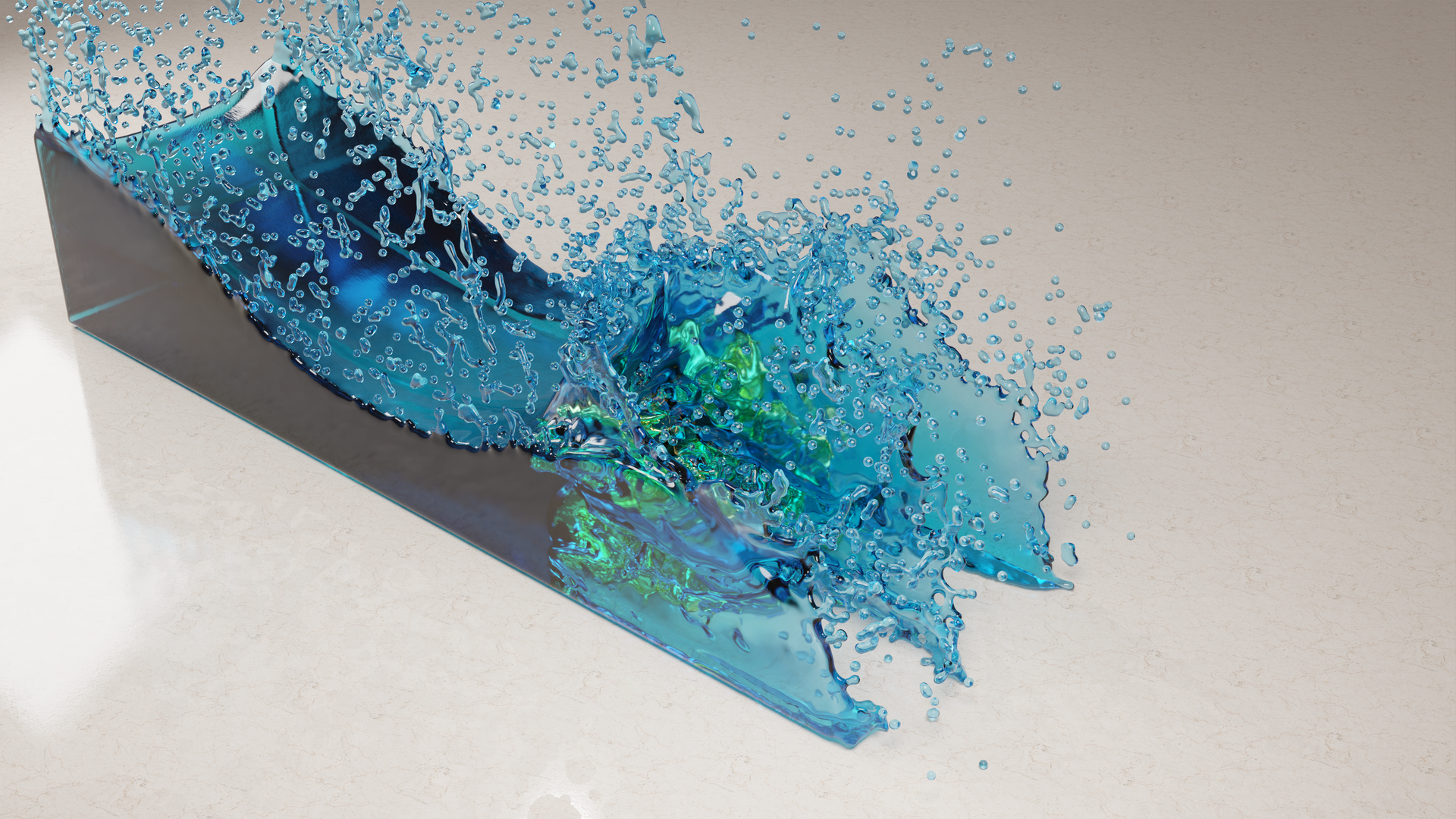} &
\end{tabular}
  \caption{Large Scale Fluid Simulation of fluid consisting of 1.23M particles.}
  \label{fig:large_scale}
  \Description{This image displays an impressive large-scale fluid simulation generated by our SPH framework. It illustrates the framework's capability to handle vast volumes of fluid, capturing the complex dynamics and interactions on a grand scale. The visualization shows the fluid's behavior under various conditions, highlighting both the expansive reach and the intricate details within the simulation. The image is a testament to the scalability and robustness of our simulation techniques, effectively demonstrating our system's capacity to model fluid dynamics in extensive environments.}
\end{figure}

\subsection{Large Scale Fluid Simulation}
Leveraging the computational might of Taichi and CUDA on A100 GPUs, our framework excels in large-scale fluid simulations, handling millions of particles with ease. This capability is showcased in Figure~\ref{fig:large_scale}, where an immense volume of water (comprising over one 1.23M particles) dramatically falls onto a ground plane, intricately splashing against two gold dragons. This scene not only highlights our framework's capability to manage a vast number of particles but also demonstrates its precision in simulating fluid-solid interactions, particularly when the particles are moving at high speeds.

\subsection{Rigid-Fluid Coupling} 
Figure~\ref{fig:coupling} displays the intricate interplay between a fluid object and nine rigid objects within a bounded environment, which is also treated as a rigid body. Our framework effectively simulates the complex interactions and forces at play between the fluid and the rigid bodies, showcasing the nuanced behavior of the fluid as it interacts with multiple solid objects in a confined space.

\subsection{High Viscosity Fluid}
Figure~\ref{fig:high_viscosity} demonstrates the ability to simulate fluid with extremely high viscosity, utilizing the viscosity coefficient \(\mu = 13000 \, \frac{\text{kg}}{\text{m} \cdot \text{s}}\). The scene reveals a slow-moving, thick fluid that realistically mimics the behavior of substances like bituman, maintaining stability and realism throughout the simulation. 

\begin{figure}[h]
  \setlength{\tabcolsep}{1pt}
\begin{tabular}{ccc}
    \centering
    &\includegraphics[width=0.96\linewidth]{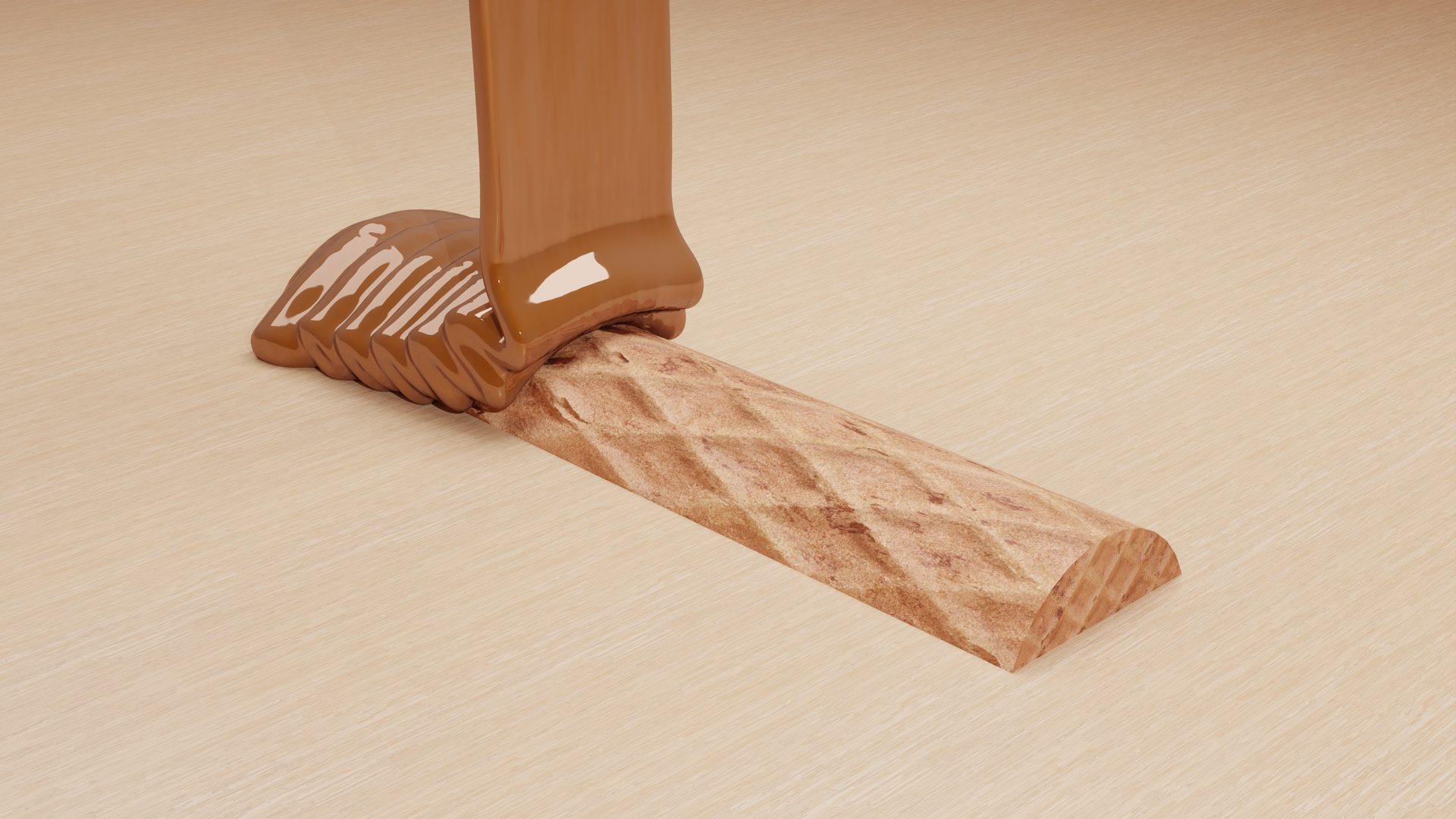} &  \\
    &\includegraphics[width=0.96\linewidth]{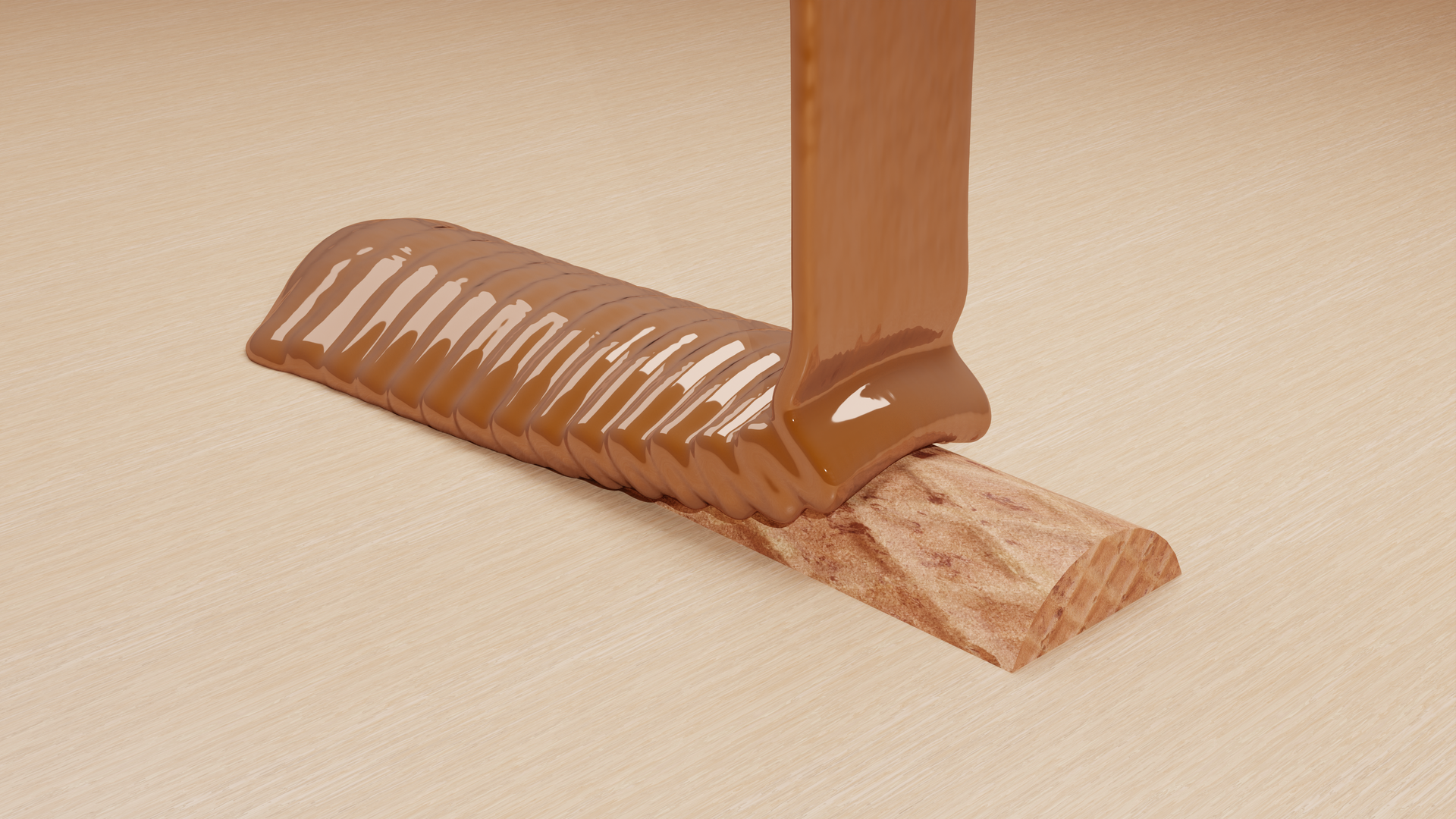} &  
\end{tabular}
  \caption{Natural buckling effect produced by our framework.}
  \label{fig:buckling}
  \Description{This figure depicts the phenomenon of the buckling effect in a fluid simulation environment. It showcases how fluid dynamics respond to external pressure or forces, leading to a characteristic buckling pattern. The visualization captures the intricate patterns and deformations that occur as a result of these forces, highlighting the fluid's behavior under stress. The image is detailed, showing the fluid’s response in a clear and vivid manner, making it an excellent representation of the buckling effect in fluid dynamics.}
\end{figure}

\subsection{Buckling Effect}
The buckling effect, commonly observed in viscous fluids, refers to the phenomenon where a fluid, under the influence of its own weight and viscosity, creates folds or wrinkles as it collapses or settles. This effect is prominently seen in materials like thick paints or molten chocolate as they are poured onto a surface.

In our simulation, as showcased in Figure~\ref{fig:buckling}, we captures this intriguing phenomenon with a simulation that resembles molten chocolate cascading over a biscuit. The simulation captures the intricate folds and layering of the fluid, creating a realistic representation of the buckling effect. The fluid's viscosity is tuned to mimic the dense, creamy texture of chocolate.  This scene  demonstrates our framework's ability to reproduce the nuanced behavior of real-world substances.

\begin{figure*}
  \centering
  \includegraphics[width=0.9\linewidth]{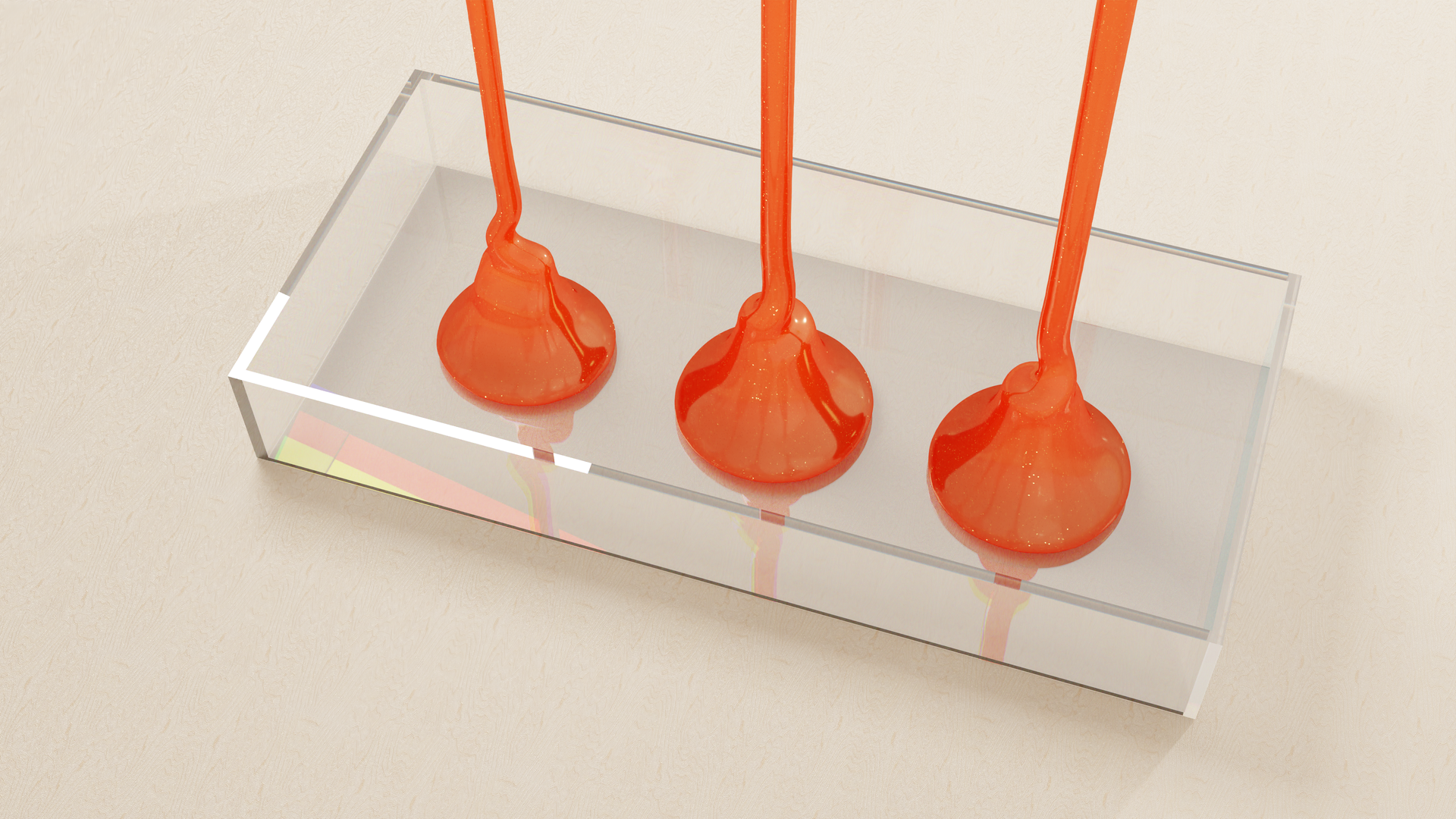}
  \caption{Simulation of coiling effect.}
  \label{fig:coiling}
  \Description{This figure presents a detailed visualization of the coiling effect as simulated in our SPH framework. It showcases the complex behavior of a fluid or semi-fluid material as it undergoes coiling, a phenomenon often observed in viscous substances under the influence of gravity. The image captures the intricate and realistic representation of the coiling patterns, highlighting the fluid's progressively looping and piling motion. The clarity and detail in the visualization emphasize the framework's capability to accurately simulate such complex fluid dynamics.}
\end{figure*}

\subsection{Coiling Effect}
The coiling effect is a fascinating fluid dynamic phenomenon often observed when a viscous fluid, such as honey or syrup, is poured onto a surface. As the fluid falls, it naturally forms a series of coils or loops, a result of the interplay between the fluid's viscosity, gravity, and the momentum of the pouring action.

Our demonstration, shown in Figure~\ref{fig:coiling}, successfully captures this intricate effect. The simulation features a fluid with a striking orange hue and a slightly glossy texture, creating an engaging visual experience. As the fluid pours down, it naturally forms coiled structures, showcasing the dynamic and complex nature of fluid behavior under the influence of gravity and viscosity. This simulation reflects the flexibility and precision of our framework in replicating such sophisticated fluid dynamics.

\begin{table*}
  \caption{Performance Comparison Across Different Computation Backends and Hardware Platforms. The first row represents the baseline configuration.}
  \label{tab:backends}
  \begin{tabular}{cccc}
    \toprule
    Computation Backend & Hardware Platform & Time Per Step [s] & Acceleration Factor \\
    \midrule
    CPU & Intel Xeon Gold 5218R, 2x20 Cores, 80 Threads & \(2.06\) & -- \\
    CPU & AMD EPYC 7713, 2x64 Cores, 256 Threads & \(0.440\) & \(4.68\) \\
    Vulkan & Nvidia GeForce RTX 3090 & \(0.111\) & \( 18.6\) \\
    CUDA & Nvidia GeForce RTX 3090 & \(0.0477\) & \( 43.2\) \\
    Vulkan & Nvidia A100 Tensor Core GPU & \(0.101\) & \( 20.4\) \\
    CUDA & Nvidia A100 Tensor Core GPU & \(0.0401\) & \(51.4\) \\

    \bottomrule
  \end{tabular}
\end{table*}

\subsection{Performance Benchmarking}
In this part, we focus on comparing the performance of different computation backends using the Taichi framework, which supports CPU, CUDA, and Vulkan. Our experiments utilize the WCSPH algorithm, chosen for its consistent computational demands, unlike PPE pressure solvers that can vary in computational rounds per step.

To optimize our testing conditions, especially for CPU computations, we increased the support radius to accelerate the prefix sum calculation. The performance metrics are based on the average time taken for a single simulation step, calculated over the initial 500 steps. To ensure fairness in our comparison, we excluded the time for prefix sum calculations from our measurements, as Taichi's prefix sum calculator does not support the CPU backend.

Our performance comparison, as shown in Table~\ref{tab:backends}, covers three computation backends across two different hardware configurations, each equipped with distinct CPUs and GPUs. The results clearly indicate that the CUDA backend consistently delivers the best performance across the same hardware platforms. Specifically, the Nvidia A100 Tensor Core GPU, when running on CUDA, demonstrates superior efficiency and speed, outperforming other combinations of hardware and computational backends.

\section{Conclusion and Future Work}
In this project, we have presented a robust SPH simulation framework designed for accurate and realistic fluid dynamics modeling. The integration of various SPH algorithms, combined with techniques for rigid-fluid interaction and high viscosity fluid simulation, demonstrates the framework's versatility and robustness. Utilizing CUDA and Taichi for computational efficiency, our framework excels in large-scale simulations. Our diverse simulations highlight the framework's capability in simulating fluid behaviors under various scenarios.

Looking ahead, we identify several key areas for further development:

\begin{itemize}
  \item \textbf{Optimization of IISPH and PBF:} Our current implementation faces challenges with Implicit Incompressible SPH (IISPH) \cite{iisph} and Position Based Fluid (PBF) \cite{pbf}. IISPH is not functioning as expected, and PBF has yet to be fitted into the unified framework. Future efforts will focus on resolving these issues to fully leverage the potential of these algorithms, enhancing the framework's versatility.

  \item \textbf{Advancements in Rigid-Body Dynamics:} Presently, our framework utilizes a "weak coupling" method for rigid-fluid interactions, assuming constant velocities for rigid bodies during fluid dynamics calculations. However, as suggested in \cite{interlinked}, adopting a "strong coupling" approach can yield more realistic results. This involves accounting for the acceleration of rigid bodies during fluid interaction, offering a more comprehensive and accurate representation of fluid-rigid dynamics. Future work will explore the integration of this "strong coupling" method to elevate the realism in our simulations.
\end{itemize}

These directions aim to expand the capability and applicability of our SPH simulation framework.

\bibliographystyle{ACM-Reference-Format}
\bibliography{sample-base}


\begin{thebibliography}{27}


\ifx \showCODEN    \undefined \def \showCODEN     #1{\unskip}     \fi
\ifx \showDOI      \undefined \def \showDOI       #1{#1}\fi
\ifx \showISBNx    \undefined \def \showISBNx     #1{\unskip}     \fi
\ifx \showISBNxiii \undefined \def \showISBNxiii  #1{\unskip}     \fi
\ifx \showISSN     \undefined \def \showISSN      #1{\unskip}     \fi
\ifx \showLCCN     \undefined \def \showLCCN      #1{\unskip}     \fi
\ifx \shownote     \undefined \def \shownote      #1{#1}          \fi
\ifx \showarticletitle \undefined \def \showarticletitle #1{#1}   \fi
\ifx \showURL      \undefined \def \showURL       {\relax}        \fi
\providecommand\bibfield[2]{#2}
\providecommand\bibinfo[2]{#2}
\providecommand\natexlab[1]{#1}
\providecommand\showeprint[2][]{arXiv:#2}

\bibitem[Adams et~al\mbox{.}(2007)]%
        {adaptive}
\bibfield{author}{\bibinfo{person}{Bart Adams}, \bibinfo{person}{Mark Pauly},
  \bibinfo{person}{Richard Keiser}, {and} \bibinfo{person}{Leonidas~J.
  Guibas}.} \bibinfo{year}{2007}\natexlab{}.
\newblock \showarticletitle{Adaptively Sampled Particle Fluids}.
\newblock \bibinfo{journal}{\emph{ACM Trans. Graph.}} \bibinfo{volume}{26},
  \bibinfo{number}{3} (\bibinfo{date}{jul} \bibinfo{year}{2007}),
  \bibinfo{pages}{48–es}.
\newblock
\showISSN{0730-0301}
\urldef\tempurl%
\url{https://doi.org/10.1145/1276377.1276437}
\showDOI{\tempurl}


\bibitem[Akinci et~al\mbox{.}(2013)]%
        {akinci2013}
\bibfield{author}{\bibinfo{person}{Nadir Akinci}, \bibinfo{person}{Jens
  Cornelis}, \bibinfo{person}{Gizem Akinci}, {and} \bibinfo{person}{Matthias
  Teschner}.} \bibinfo{year}{2013}\natexlab{}.
\newblock \showarticletitle{Coupling elastic solids with smoothed particle
  hydrodynamics fluids}.
\newblock \bibinfo{journal}{\emph{Computer Animation and Virtual Worlds}}
  \bibinfo{volume}{24}, \bibinfo{number}{3-4} (\bibinfo{year}{2013}),
  \bibinfo{pages}{195--203}.
\newblock
\urldef\tempurl%
\url{https://doi.org/10.1002/cav.1499}
\showDOI{\tempurl}
\showeprint{https://onlinelibrary.wiley.com/doi/pdf/10.1002/cav.1499}


\bibitem[Akinci et~al\mbox{.}(2012)]%
        {akinci2012}
\bibfield{author}{\bibinfo{person}{Nadir Akinci}, \bibinfo{person}{Markus
  Ihmsen}, \bibinfo{person}{Gizem Akinci}, \bibinfo{person}{Barbara
  Solenthaler}, {and} \bibinfo{person}{Matthias Teschner}.}
  \bibinfo{year}{2012}\natexlab{}.
\newblock \showarticletitle{Versatile Rigid-Fluid Coupling for Incompressible
  SPH}.
\newblock \bibinfo{journal}{\emph{ACM Trans. Graph.}} \bibinfo{volume}{31},
  \bibinfo{number}{4}, Article \bibinfo{articleno}{62} (\bibinfo{date}{jul}
  \bibinfo{year}{2012}), \bibinfo{numpages}{8}~pages.
\newblock
\showISSN{0730-0301}
\urldef\tempurl%
\url{https://doi.org/10.1145/2185520.2185558}
\showDOI{\tempurl}


\bibitem[Becker and Teschner(2007)]%
        {wcsph}
\bibfield{author}{\bibinfo{person}{Markus Becker} {and}
  \bibinfo{person}{Matthias Teschner}.} \bibinfo{year}{2007}\natexlab{}.
\newblock \showarticletitle{Weakly Compressible SPH for Free Surface Flows}
  \emph{(\bibinfo{series}{SCA '07})}. \bibinfo{publisher}{Eurographics
  Association}, \bibinfo{address}{Goslar, DEU}.
\newblock
\showISBNx{9781595936240}


\bibitem[Bender and Koschier(2015)]%
        {dfsph}
\bibfield{author}{\bibinfo{person}{Jan Bender} {and} \bibinfo{person}{Dan
  Koschier}.} \bibinfo{year}{2015}\natexlab{}.
\newblock \showarticletitle{Divergence-Free Smoothed Particle Hydrodynamics}.
  In \bibinfo{booktitle}{\emph{Proceedings of the 14th ACM SIGGRAPH /
  Eurographics Symposium on Computer Animation}} (Los Angeles, California)
  \emph{(\bibinfo{series}{SCA '15})}. \bibinfo{publisher}{Association for
  Computing Machinery}, \bibinfo{address}{New York, NY, USA},
  \bibinfo{pages}{147–155}.
\newblock
\showISBNx{9781450334969}
\urldef\tempurl%
\url{https://doi.org/10.1145/2786784.2786796}
\showDOI{\tempurl}


\bibitem[Bender and Koschier(2017)]%
        {bk17}
\bibfield{author}{\bibinfo{person}{Jan Bender} {and} \bibinfo{person}{Dan
  Koschier}.} \bibinfo{year}{2017}\natexlab{}.
\newblock \showarticletitle{Divergence-Free SPH for Incompressible and Viscous
  Fluids}.
\newblock  \bibinfo{volume}{23}, \bibinfo{number}{3} (\bibinfo{date}{mar}
  \bibinfo{year}{2017}), \bibinfo{pages}{1193–1206}.
\newblock
\showISSN{1077-2626}
\urldef\tempurl%
\url{https://doi.org/10.1109/TVCG.2016.2578335}
\showDOI{\tempurl}


\bibitem[Bender et~al\mbox{.}(2019)]%
        {volume2019}
\bibfield{author}{\bibinfo{person}{Jan Bender}, \bibinfo{person}{Tassilo
  Kugelstadt}, \bibinfo{person}{Marcel Weiler}, {and} \bibinfo{person}{Dan
  Koschier}.} \bibinfo{year}{2019}\natexlab{}.
\newblock \showarticletitle{Volume Maps: An Implicit Boundary Representation
  for SPH}. In \bibinfo{booktitle}{\emph{Proceedings of the 12th ACM SIGGRAPH
  Conference on Motion, Interaction and Games}} (Newcastle upon Tyne, United
  Kingdom) \emph{(\bibinfo{series}{MIG '19})}. \bibinfo{publisher}{Association
  for Computing Machinery}, \bibinfo{address}{New York, NY, USA}, Article
  \bibinfo{articleno}{26}, \bibinfo{numpages}{10}~pages.
\newblock
\showISBNx{9781450369947}
\urldef\tempurl%
\url{https://doi.org/10.1145/3359566.3360077}
\showDOI{\tempurl}


\bibitem[Bodin et~al\mbox{.}(2012)]%
        {cf}
\bibfield{author}{\bibinfo{person}{Kenneth Bodin}, \bibinfo{person}{Claude
  Lacoursiere}, {and} \bibinfo{person}{Martin Servin}.}
  \bibinfo{year}{2012}\natexlab{}.
\newblock \showarticletitle{Constraint Fluids}.
\newblock \bibinfo{journal}{\emph{IEEE Transactions on Visualization and
  Computer Graphics}} \bibinfo{volume}{18}, \bibinfo{number}{3}
  (\bibinfo{date}{mar} \bibinfo{year}{2012}), \bibinfo{pages}{516–526}.
\newblock
\showISSN{1077-2626}
\urldef\tempurl%
\url{https://doi.org/10.1109/TVCG.2011.29}
\showDOI{\tempurl}


\bibitem[Gissler et~al\mbox{.}(2019)]%
        {interlinked}
\bibfield{author}{\bibinfo{person}{Christoph Gissler}, \bibinfo{person}{Andreas
  Peer}, \bibinfo{person}{Stefan Band}, \bibinfo{person}{Jan Bender}, {and}
  \bibinfo{person}{Matthias Teschner}.} \bibinfo{year}{2019}\natexlab{}.
\newblock \showarticletitle{Interlinked SPH Pressure Solvers for Strong
  Fluid-Rigid Coupling}.
\newblock \bibinfo{journal}{\emph{ACM Trans. Graph.}} \bibinfo{volume}{38},
  \bibinfo{number}{1}, Article \bibinfo{articleno}{5} (\bibinfo{date}{jan}
  \bibinfo{year}{2019}), \bibinfo{numpages}{13}~pages.
\newblock
\showISSN{0730-0301}
\urldef\tempurl%
\url{https://doi.org/10.1145/3284980}
\showDOI{\tempurl}


\bibitem[Greff et~al\mbox{.}(2022)]%
        {bullet}
\bibfield{author}{\bibinfo{person}{Klaus Greff}, \bibinfo{person}{Francois
  Belletti}, \bibinfo{person}{Lucas Beyer}, \bibinfo{person}{Carl Doersch},
  \bibinfo{person}{Yilun Du}, \bibinfo{person}{Daniel Duckworth},
  \bibinfo{person}{David~J Fleet}, \bibinfo{person}{Dan Gnanapragasam},
  \bibinfo{person}{Florian Golemo}, \bibinfo{person}{Charles Herrmann},
  \bibinfo{person}{Thomas Kipf}, \bibinfo{person}{Abhijit Kundu},
  \bibinfo{person}{Dmitry Lagun}, \bibinfo{person}{Issam Laradji},
  \bibinfo{person}{Hsueh-Ti~(Derek) Liu}, \bibinfo{person}{Henning Meyer},
  \bibinfo{person}{Yishu Miao}, \bibinfo{person}{Derek Nowrouzezahrai},
  \bibinfo{person}{Cengiz Oztireli}, \bibinfo{person}{Etienne Pot},
  \bibinfo{person}{Noha Radwan}, \bibinfo{person}{Daniel Rebain},
  \bibinfo{person}{Sara Sabour}, \bibinfo{person}{Mehdi S.~M. Sajjadi},
  \bibinfo{person}{Matan Sela}, \bibinfo{person}{Vincent Sitzmann},
  \bibinfo{person}{Austin Stone}, \bibinfo{person}{Deqing Sun},
  \bibinfo{person}{Suhani Vora}, \bibinfo{person}{Ziyu Wang},
  \bibinfo{person}{Tianhao Wu}, \bibinfo{person}{Kwang~Moo Yi},
  \bibinfo{person}{Fangcheng Zhong}, {and} \bibinfo{person}{Andrea
  Tagliasacchi}.} \bibinfo{year}{2022}\natexlab{}.
\newblock \showarticletitle{Kubric: a scalable dataset generator}.
\newblock  (\bibinfo{year}{2022}).
\newblock


\bibitem[Hu et~al\mbox{.}(2020)]%
        {hu2019difftaichi}
\bibfield{author}{\bibinfo{person}{Yuanming Hu}, \bibinfo{person}{Luke
  Anderson}, \bibinfo{person}{Tzu-Mao Li}, \bibinfo{person}{Qi Sun},
  \bibinfo{person}{Nathan Carr}, \bibinfo{person}{Jonathan Ragan-Kelley}, {and}
  \bibinfo{person}{Fr{\'e}do Durand}.} \bibinfo{year}{2020}\natexlab{}.
\newblock \showarticletitle{DiffTaichi: Differentiable Programming for Physical
  Simulation}.
\newblock \bibinfo{journal}{\emph{ICLR}} (\bibinfo{year}{2020}).
\newblock


\bibitem[Hu et~al\mbox{.}(2019)]%
        {hu2019taichi}
\bibfield{author}{\bibinfo{person}{Yuanming Hu}, \bibinfo{person}{Tzu-Mao Li},
  \bibinfo{person}{Luke Anderson}, \bibinfo{person}{Jonathan Ragan-Kelley},
  {and} \bibinfo{person}{Fr{\'e}do Durand}.} \bibinfo{year}{2019}\natexlab{}.
\newblock \showarticletitle{Taichi: a language for high-performance computation
  on spatially sparse data structures}.
\newblock \bibinfo{journal}{\emph{ACM Transactions on Graphics (TOG)}}
  \bibinfo{volume}{38}, \bibinfo{number}{6} (\bibinfo{year}{2019}),
  \bibinfo{pages}{201}.
\newblock


\bibitem[Hu et~al\mbox{.}(2021)]%
        {hu2021quantaichi}
\bibfield{author}{\bibinfo{person}{Yuanming Hu}, \bibinfo{person}{Jiafeng Liu},
  \bibinfo{person}{Xuanda Yang}, \bibinfo{person}{Mingkuan Xu},
  \bibinfo{person}{Ye Kuang}, \bibinfo{person}{Weiwei Xu},
  \bibinfo{person}{Qiang Dai}, \bibinfo{person}{William~T. Freeman}, {and}
  \bibinfo{person}{Frédo Durand}.} \bibinfo{year}{2021}\natexlab{}.
\newblock \showarticletitle{QuanTaichi: A Compiler for Quantized Simulations}.
\newblock \bibinfo{journal}{\emph{ACM Transactions on Graphics (TOG)}}
  \bibinfo{volume}{40}, \bibinfo{number}{4} (\bibinfo{year}{2021}).
\newblock


\bibitem[Ihmsen et~al\mbox{.}(2014)]%
        {iisph}
\bibfield{author}{\bibinfo{person}{Markus Ihmsen}, \bibinfo{person}{Jens
  Cornelis}, \bibinfo{person}{Barbara Solenthaler},
  \bibinfo{person}{Christopher Horvath}, {and} \bibinfo{person}{Matthias
  Teschner}.} \bibinfo{year}{2014}\natexlab{}.
\newblock \showarticletitle{Implicit Incompressible SPH}.
\newblock  \bibinfo{volume}{20}, \bibinfo{number}{3} (\bibinfo{date}{mar}
  \bibinfo{year}{2014}), \bibinfo{pages}{426–435}.
\newblock
\showISSN{1077-2626}
\urldef\tempurl%
\url{https://doi.org/10.1109/TVCG.2013.105}
\showDOI{\tempurl}


\bibitem[Koschier and Bender(2017)]%
        {density_map}
\bibfield{author}{\bibinfo{person}{Dan Koschier} {and} \bibinfo{person}{Jan
  Bender}.} \bibinfo{year}{2017}\natexlab{}.
\newblock \showarticletitle{Density Maps for Improved SPH Boundary Handling}.
  In \bibinfo{booktitle}{\emph{Proceedings of the ACM SIGGRAPH / Eurographics
  Symposium on Computer Animation}} (Los Angeles, California)
  \emph{(\bibinfo{series}{SCA '17})}. \bibinfo{publisher}{Association for
  Computing Machinery}, \bibinfo{address}{New York, NY, USA}, Article
  \bibinfo{articleno}{1}, \bibinfo{numpages}{10}~pages.
\newblock
\showISBNx{9781450350914}
\urldef\tempurl%
\url{https://doi.org/10.1145/3099564.3099565}
\showDOI{\tempurl}


\bibitem[Koschier et~al\mbox{.}(2019)]%
        {tutorial2019}
\bibfield{author}{\bibinfo{person}{Dan Koschier}, \bibinfo{person}{Jan Bender},
  \bibinfo{person}{Barbara Solenthaler}, {and} \bibinfo{person}{Matthias
  Teschner}.} \bibinfo{year}{2019}\natexlab{}.
\newblock \showarticletitle{{Smoothed Particle Hydrodynamics Techniques for the
  Physics Based Simulation of Fluids and Solids}}. In
  \bibinfo{booktitle}{\emph{Eurographics 2019 - Tutorials}},
  \bibfield{editor}{\bibinfo{person}{Wenzel Jakob} {and}
  \bibinfo{person}{Enrico Puppo}} (Eds.). \bibinfo{publisher}{The Eurographics
  Association}, \bibinfo{pages}{1--41}.
\newblock
\urldef\tempurl%
\url{https://doi.org/10.2312/egt.20191035}
\showDOI{\tempurl}


\bibitem[Koschier et~al\mbox{.}(2022)]%
        {tutorial2022}
\bibfield{author}{\bibinfo{person}{Dan Koschier}, \bibinfo{person}{Jan Bender},
  \bibinfo{person}{Barbara Solenthaler}, {and} \bibinfo{person}{Matthias
  Teschner}.} \bibinfo{year}{2022}\natexlab{}.
\newblock \showarticletitle{{A Survey on SPH Methods in Computer Graphics}}.
\newblock \bibinfo{journal}{\emph{Computer Graphics Forum}}
  (\bibinfo{year}{2022}).
\newblock
\showISSN{1467-8659}
\urldef\tempurl%
\url{https://doi.org/10.1111/cgf.14508}
\showDOI{\tempurl}


\bibitem[Löschner et~al\mbox{.}(2023)]%
        {splashsurf}
\bibfield{author}{\bibinfo{person}{Fabian Löschner}, \bibinfo{person}{Timna
  Böttcher}, \bibinfo{person}{Stefan Rhys~Jeske}, {and} \bibinfo{person}{Jan
  Bender}.} \bibinfo{year}{2023}\natexlab{}.
\newblock \showarticletitle{{Weighted Laplacian Smoothing for Surface
  Reconstruction of Particle-based Fluids}}. In
  \bibinfo{booktitle}{\emph{Vision, Modeling, and Visualization}}.
  \bibinfo{publisher}{The Eurographics Association}.
\newblock
\urldef\tempurl%
\url{https://doi.org/10.2312/vmv.20231245}
\showDOI{\tempurl}


\bibitem[Macklin and M\"{u}ller(2013)]%
        {pbf}
\bibfield{author}{\bibinfo{person}{Miles Macklin} {and}
  \bibinfo{person}{Matthias M\"{u}ller}.} \bibinfo{year}{2013}\natexlab{}.
\newblock \showarticletitle{Position Based Fluids}.
\newblock  \bibinfo{volume}{32}, \bibinfo{number}{4}, Article
  \bibinfo{articleno}{104} (\bibinfo{date}{jul} \bibinfo{year}{2013}),
  \bibinfo{numpages}{12}~pages.
\newblock
\showISSN{0730-0301}
\urldef\tempurl%
\url{https://doi.org/10.1145/2461912.2461984}
\showDOI{\tempurl}


\bibitem[Monaghan(2005)]%
        {mon05}
\bibfield{author}{\bibinfo{person}{J~J Monaghan}.}
  \bibinfo{year}{2005}\natexlab{}.
\newblock \showarticletitle{Smoothed particle hydrodynamics}.
\newblock \bibinfo{journal}{\emph{Reports on Progress in Physics}}
  \bibinfo{volume}{68}, \bibinfo{number}{8} (\bibinfo{date}{jul}
  \bibinfo{year}{2005}), \bibinfo{pages}{1703}.
\newblock
\urldef\tempurl%
\url{https://doi.org/10.1088/0034-4885/68/8/R01}
\showDOI{\tempurl}


\bibitem[M\"{u}ller et~al\mbox{.}(2003)]%
        {eos_first}
\bibfield{author}{\bibinfo{person}{Matthias M\"{u}ller}, \bibinfo{person}{David
  Charypar}, {and} \bibinfo{person}{Markus Gross}.}
  \bibinfo{year}{2003}\natexlab{}.
\newblock \showarticletitle{Particle-Based Fluid Simulation for Interactive
  Applications} \emph{(\bibinfo{series}{SCA '03})}.
  \bibinfo{publisher}{Eurographics Association}, \bibinfo{address}{Goslar,
  DEU}, \bibinfo{pages}{154–159}.
\newblock
\showISBNx{1581136595}


\bibitem[M\"{u}ller et~al\mbox{.}(2007)]%
        {pbd}
\bibfield{author}{\bibinfo{person}{Matthias M\"{u}ller}, \bibinfo{person}{Bruno
  Heidelberger}, \bibinfo{person}{Marcus Hennix}, {and} \bibinfo{person}{John
  Ratcliff}.} \bibinfo{year}{2007}\natexlab{}.
\newblock \showarticletitle{Position Based Dynamics}.
\newblock \bibinfo{journal}{\emph{J. Vis. Comun. Image Represent.}}
  \bibinfo{volume}{18}, \bibinfo{number}{2} (\bibinfo{date}{apr}
  \bibinfo{year}{2007}), \bibinfo{pages}{109–118}.
\newblock
\showISSN{1047-3203}
\urldef\tempurl%
\url{https://doi.org/10.1016/j.jvcir.2007.01.005}
\showDOI{\tempurl}


\bibitem[Peer et~al\mbox{.}(2015)]%
        {pict15}
\bibfield{author}{\bibinfo{person}{Andreas Peer}, \bibinfo{person}{Markus
  Ihmsen}, \bibinfo{person}{Jens Cornelis}, {and} \bibinfo{person}{Matthias
  Teschner}.} \bibinfo{year}{2015}\natexlab{}.
\newblock \showarticletitle{An Implicit Viscosity Formulation for SPH Fluids}.
\newblock \bibinfo{journal}{\emph{ACM Trans. Graph.}} \bibinfo{volume}{34},
  \bibinfo{number}{4}, Article \bibinfo{articleno}{114} (\bibinfo{date}{jul}
  \bibinfo{year}{2015}), \bibinfo{numpages}{10}~pages.
\newblock
\showISSN{0730-0301}
\urldef\tempurl%
\url{https://doi.org/10.1145/2766925}
\showDOI{\tempurl}


\bibitem[Peer and Teschner(2017)]%
        {pt16}
\bibfield{author}{\bibinfo{person}{Andreas Peer} {and}
  \bibinfo{person}{Matthias Teschner}.} \bibinfo{year}{2017}\natexlab{}.
\newblock \showarticletitle{Prescribed Velocity Gradients for Highly Viscous
  SPH Fluids with Vorticity Diffusion}.
\newblock \bibinfo{journal}{\emph{IEEE Transactions on Visualization and
  Computer Graphics}} \bibinfo{volume}{23}, \bibinfo{number}{12}
  (\bibinfo{year}{2017}), \bibinfo{pages}{2656--2662}.
\newblock
\urldef\tempurl%
\url{https://doi.org/10.1109/TVCG.2016.2636144}
\showDOI{\tempurl}


\bibitem[Solenthaler and Pajarola(2009)]%
        {pcisph}
\bibfield{author}{\bibinfo{person}{B. Solenthaler} {and} \bibinfo{person}{R.
  Pajarola}.} \bibinfo{year}{2009}\natexlab{}.
\newblock \showarticletitle{Predictive-Corrective Incompressible SPH}. In
  \bibinfo{booktitle}{\emph{ACM SIGGRAPH 2009 Papers}} (New Orleans, Louisiana)
  \emph{(\bibinfo{series}{SIGGRAPH '09})}. \bibinfo{publisher}{Association for
  Computing Machinery}, \bibinfo{address}{New York, NY, USA}, Article
  \bibinfo{articleno}{40}, \bibinfo{numpages}{6}~pages.
\newblock
\showISBNx{9781605587264}
\urldef\tempurl%
\url{https://doi.org/10.1145/1576246.1531346}
\showDOI{\tempurl}


\bibitem[Takahashi et~al\mbox{.}(2015)]%
        {tdf15}
\bibfield{author}{\bibinfo{person}{Tetsuya Takahashi},
  \bibinfo{person}{Yoshinori Dobashi}, \bibinfo{person}{Issei Fujishiro},
  \bibinfo{person}{Tomoyuki Nishita}, {and} \bibinfo{person}{Ming~C. Lin}.}
  \bibinfo{year}{2015}\natexlab{}.
\newblock \showarticletitle{Implicit Formulation for SPH-based Viscous Fluids}.
\newblock \bibinfo{journal}{\emph{Computer Graphics Forum}}
  \bibinfo{volume}{34}, \bibinfo{number}{2} (\bibinfo{year}{2015}),
  \bibinfo{pages}{493--502}.
\newblock
\urldef\tempurl%
\url{https://doi.org/10.1111/cgf.12578}
\showDOI{\tempurl}
\showeprint{https://onlinelibrary.wiley.com/doi/pdf/10.1111/cgf.12578}


\bibitem[Weiler et~al\mbox{.}(2018)]%
        {wkbb18}
\bibfield{author}{\bibinfo{person}{Marcel Weiler}, \bibinfo{person}{Dan
  Koschier}, \bibinfo{person}{Magnus Brand}, {and} \bibinfo{person}{Jan
  Bender}.} \bibinfo{year}{2018}\natexlab{}.
\newblock \showarticletitle{A Physically Consistent Implicit Viscosity Solver
  for SPH Fluids}.
\newblock \bibinfo{journal}{\emph{Computer Graphics Forum}}
  \bibinfo{volume}{37}, \bibinfo{number}{2} (\bibinfo{year}{2018}),
  \bibinfo{pages}{145--155}.
\newblock
\urldef\tempurl%
\url{https://doi.org/10.1111/cgf.13349}
\showDOI{\tempurl}
\showeprint{https://onlinelibrary.wiley.com/doi/pdf/10.1111/cgf.13349}


\end{thebibliography}




\end{document}